\documentclass[reprint,NumberedRefs,floatfix,TurnOnLineNumbers]{JASAnew}%NumberedRefs
\usepackage{url}

\usepackage{jabbrv}
\usepackage{multirow}

\usepackage[sort&compress]{natbib}

\usepackage{lmodern}
\usepackage{xcolor}

\begin{document}

\title[AcousticsML]{Machine Learning in Acoustics: A Review and Open-Source Repository}
\author{Ryan A. McCarthy\textsuperscript{*}}
\email{r1mccarthy@ucsd.edu}
\affiliation{Scripps Institution of Oceanography, UC San Diego, La Jolla, CA 92093, USA}
\author{You Zhang}

\affiliation{Electrical and Computer Engineering, University of Rochester, Rochester, NY 14627, USA}
\author{Samuel A. Verburg}

\affiliation{Electrical and Photonics Engineering, Technical University of Denmark, 2800 Kgs.\ Lyngby, Denmark}
\author{William F. Jenkins}

\affiliation{Scripps Institution of Oceanography, UC San Diego, La Jolla, CA 92093, USA}
\author{Peter Gerstoft}

\affiliation{Scripps Institution of Oceanography, UC San Diego, La Jolla, CA 92093, USA}
\affiliation{Electrical and Photonics Engineering, Technical University of Denmark, 2800 Kgs.\ Lyngby, Denmark}

\preprint{Author, JASA}

\begin{abstract}
% Acoustic data provide scientific and engineering insights in fields ranging from biology and communications to ocean and earth sciences. We survey recent advances and the transformative potential of machine learning (ML) in acoustics, including deep learning (DL). Using the Python high-level programming language, we demonstrate a broad collection of ML techniques, often based on statistics, to detect and find patterns in data automatically. We have ML examples from ocean acoustics, room acoustics, and personalized spatial audio for personalized head-related transfer functions modeling in gaming. The tutorial includes a set of Jupyter notebook examples on GitHub demonstrating ML benefits. 
Acoustic data provide scientific and engineering insights in fields ranging from bioacoustics and communications to ocean and earth sciences. In this review, we survey recent advances and the transformative potential of machine learning (ML) in acoustics, including deep learning (DL). Using the Python high-level programming language, we demonstrate a broad collection of ML techniques to detect and find patterns for classification, regression, and generation in acoustics data automatically. We have ML examples including acoustic data classification, generative modeling for spatial audio, and physics-informed neural networks. This work includes AcousticsML, a set of practical Jupyter notebook examples on GitHub demonstrating ML benefits and encouraging researchers and practitioners to apply reproducible data-driven approaches to acoustic challenges.
\end{abstract}

\maketitle

\section{Introduction}
\begin{figure}
\centering
\includegraphics[width=0.8\linewidth]{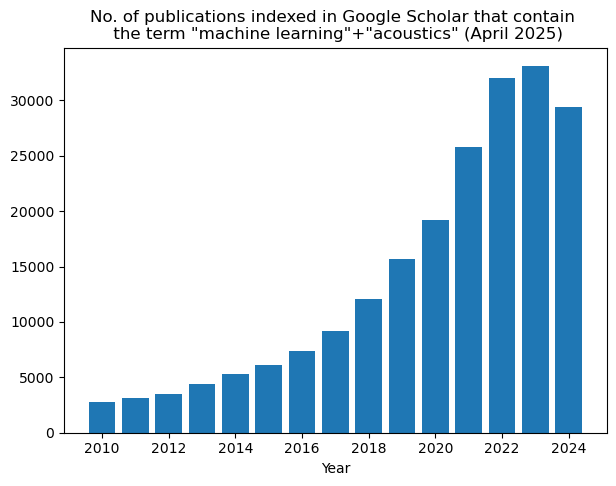}
\caption{Number of publications containing the keywords ``machine learning" and ``acoustics", using the script in \onlinecite{Volker2018pold}.}
\label{fig:numberpublications}
\end{figure}

Machine learning (ML)\cite{bishop2006pattern,theodoridis2015machine,goodfellow2016deep,murphy2022probabilistic} has become a valuable tool for processing and analyzing large acoustic datasets and improving the interpretability of acoustic data.\cite{bianco2019machine,cai2021physics,karniadakis2021physics,michalopoulou2021introduction,grumiaux2022survey,cunha2023review,niu2023advances,stanczyk2014feature}
A search of publications containing ``machine learning'' and ``acoustics'' as keywords reveals that interest in applying ML solutions to problems in acoustics has grown exponentially in the past 15 years, as illustrated by Fig.~\ref{fig:numberpublications}.
This interest is partly driven by technological advances in acoustics, which produce increasingly large datasets. These datasets present new challenges for evaluation and interpretation, as the time and resources required to analyze such data manually are becoming prohibitively expensive.
ML algorithms offer solutions to some of these challenges with their ability to identify patterns and trends through statistical and non-linear approaches, offering fast, efficient, and reproducible approaches that scale to the growing size and complexity of acoustic datasets.

Although ML provides powerful tools for acoustic data processing and analysis, applying these techniques to acoustics remains difficult without guides that address the field's unique data challenges.
Many fields (e.g., seismology, econometrics, meteorology) have tutorials demonstrating domain-specific ML applications, but establishing direct parallels with acoustic applications is often not straightforward.
This work addresses these challenges by providing an open-source GitHub repository, designated as AcousticsML, featuring Jupyter Notebooks with diverse ML applications in acoustics, including time series analysis, physics-based modeling, classification, clustering, and techniques for managing large datasets.
Additionally, we describe several structured workflows for applying machine learning to acoustic data, with each pipeline tailored to specific applications while offering a replicable framework.
Although ML libraries are available in many programming languages and software suites, we rely on Python since it provides researchers with a mature ecosystem for ML development, offering comprehensive libraries, extensive documentation, and readable syntax.

Using ML is an inherently data-intensive task.
The curation of publicly available data sets that can be used to evaluate the suitability of ML models for certain types of data has aided ML development.
However, the field of acoustics has fewer defined baseline datasets with which to train and evaluate models.
In this work we highlight several publicly available acoustic datasets for speech recognition\cite{Wichern2019WHAM, Maciejewski2020WHAMR}, ambient noise and sound classification \cite{stowell2013open, Xiao2024WildDESED,sayigh2016watkins,gemmeke2017audio,becker2024audiomnist}, room impulse responses~\cite{szoke2019building,MeshRIR,Verburg2024room}, and head-related transfer functions \cite{brinkmann2019cross, ghorbal2020computed, guezenoc2020wide, engel2023sonicom}; additional data sets can be found in online repositories such as Kaggle or Zenodo.
Additionally, we show models developed with smaller or simulated datasets that are nevertheless able to provide adequate information features on which the models can train.
Though simulations offer cost-efficient training alternatives when real data are limited, potential biases and realism limitations must be carefully addressed, a challenge that presents both constraints and opportunities for innovation in acoustic ML.

\section{Recent Advances in ML Techniques}
In Ref.~\onlinecite{bianco2019machine}, several ML algorithms were highlighted to provide relevant details about acoustic applications. In particular, ML principles, supervised and unsupervised, and deep learning (DL) theory were discussed in detail to demonstrate key advancements in ML and how they can be applied to acoustic datasets. Models are trained from measurements and are formulated as
\begin{equation}
    y = f(x) +\epsilon,
\end{equation}
where $y$ is the output, $x$ is a single input (observation) with N features, $f(x)$ is the model that maps input features to the output, and $\epsilon$ is the uncertainty in model prediction. The trained model algorithms can vary from linear to non-linear estimators that best fit any application. Although many ML techniques have been covered in Ref.~\onlinecite{bianco2019machine}, this paper briefly reviews recent ML techniques, emphasizing acoustic applications through open-source code. 

\subsection{Generative Models}
Generative models have become essential tools in acoustic ML. As DL technology progresses, three main types of generative models have emerged at the forefront: Generative Adversarial Networks (GANs)\cite{fernandez2023generative}, Variational Autoencoders (VAEs)\cite{jenkins2021unsupervised,bianco2021semi,saha2025leveraging}, and Diffusion Models\cite{yang2023diffsound,li2024learning}.

\begin{figure*}
    \centering
    \includegraphics[width=0.98\linewidth]{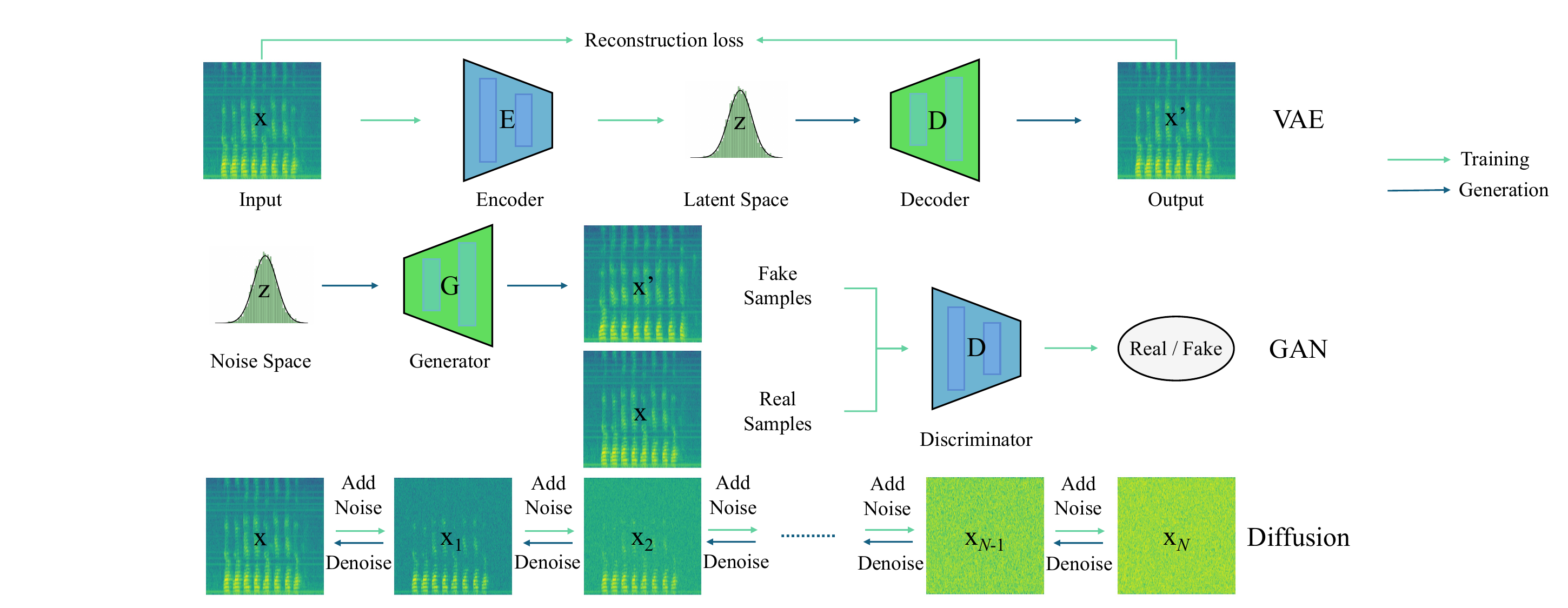}
\caption{Illustration of different types of generative models with generating audio spectrograms as an example. 1) Variational Autoencoder (VAE): Trained to reconstruct the spectrogram while regularizing the latent space to a Gaussian distribution. The decoder (D) generates spectrograms by decoding Gaussian noise. 2) Generative Adversarial Networks (GANs): The generator (G) synthesizes spectrograms from random noise, while the discriminator (D) distinguishes real from generated samples in an adversarial training setup. 3) Diffusion Models: Trained by gradually adding noise to the spectrogram, then learning to reverse the process using neural networks. After training, the model generates spectrograms by reversing the noise process from Gaussian noise.
    \label{fig:generative}}
\end{figure*}

\subsubsection{Variational Autoencoder}
\label{ssec:vae}

VAEs\cite{kingma2013auto}, illustrated at the top of Fig.~\ref{fig:generative}, are generative models based on an encoder-decoder architecture, generalized from deterministic autoencoders, which are mainly used for dimensionality reduction (see Sec.~\ref{dimred_aec}). The encoder (E) maps the input data $\mathbf{x}$ to a probabilistic latent space, defined by a distribution over low-dimensional latent variables \( \mathbf{z} \), while the decoder (D) reconstructs the data from samples of this distribution. VAEs minimize the reconstruction error while ensuring that the latent space adheres to a prior distribution, typically a standard normal distribution.

The training of VAEs involves two objectives: reconstruction and regularization of the latent space. By minimizing the loss of reconstruction, the model aims to recreate the original data from its compressed version as accurately as possible. This ensures that the VAE learns an accurate representation of the data that can be used to generate new samples. Additionally, VAEs ensure that the latent space follows a simple and well-organized structure, typically similar to random noise. This step enables the model to generate diverse and realistic new samples, thereby preventing it from overfitting the training data.

Mathematically, this optimization is grounded in maximizing the variational lower bound, which can be thought of as a means to strike the best balance between fitting the model to the data and maintaining the model's simplicity by Kullback–Leibler (KL) divergence regularization. The model learns two distributions: the encoder distribution $p(\mathbf{x} | \mathbf{z})$ and the decoder distribution $q(\mathbf{z}|\mathbf{x})$.

\begin{equation}
    \log p(\mathbf{x}) \geq \mathbb{E}_{q(\mathbf{z}|\mathbf{x})} \left[ \log p(\mathbf{x} | \mathbf{z}) \right] - D_{\textit{KL}} \left( q(\mathbf{z}|\mathbf{x}) || p(\mathbf{z}) \right),
\end{equation}
where $p(\mathbf{x} | \mathbf{z})$ is the likelihood of the data given the latent variables. $p(\mathbf{z})$ is the prior distribution on the latent variables, often chosen to be a standard Gaussian $\mathcal{N}(0, I)$. $q(\mathbf{z} | \mathbf{x})$ is the approximate posterior parameterized by a neural network.

The first term encourages the decoder to produce data reconstruction close to the input, while the second term, KL divergence, penalizes the model if its latent representations deviate too much from a predefined, simple distribution (often a Gaussian), ensuring that the latent space does not overfit specific data points in the datasets.

The combination of these two objectives strikes a balance between accurate data reconstruction and a smooth, well-structured, and meaningful latent space, enabling VAEs to generate meaningful new data. For example, once trained, a VAE can generate entirely new audio samples that are similar to the training data but not identical, making it particularly useful for applications such as speech synthesis~\cite{akuzawa18expressive} or music generation~\cite{roberts2018hierarchical}, where diversity and smooth latent interpolations are crucial.

\subsubsection{Generative Adversarial Networks}
GANs\cite{goodfellow2014generative} are a class of ML models~\cite{goodfellow2020generative} that consist of two neural networks that work together: a generator (G) and a discriminator (D). These networks are trained through an adversarial process, where the generator attempts to create synthetic data samples that follow real data distribution, and the discriminator’s task is to distinguish between real and generated samples. This process is illustrated in Fig.~\ref{fig:generative} middle.

The generator takes a random noise vector \( \mathbf{z} \) as input and learns to map it to the input data space, generating samples that resemble the real data. On the other hand, the discriminator attempts to correctly classify real and generated data, outputting the probability that a given sample is real (rather than fake). The discriminator tries to minimize the objective
\begin{equation}
\min_D \mathbb{E}_{x \sim p_{\text{data}}(x)}[\log D(x)] + \mathbb{E}_{z \sim p_z(z)}[\log(1 - D(G(z)))],
\end{equation}
while the generator’s objective is to maximize the probability of the discriminator making an error in distinguishing real from fake data.
\begin{equation}
    \min_G \max_D \mathbb{E}_{x \sim p_{\text{data}}(x)}[\log D(x)] + \mathbb{E}_{z \sim p_z(z)}[\log(1 - D(G(z)))],
\end{equation}
where \( x \) represents real data, \( z \) is the random noise input, and \( G(z) \) is the synthetic sample generated by the generator.

The training proceeds in a minimax game, where the generator and discriminator continuously improve, with the generator trying to produce increasingly realistic samples and the discriminator learning to distinguish between real and fake data. Through this adversarial process, GANs can generate highly realistic data in various domains, such as images, audio, and video, making them a cornerstone of modern generative modeling.

This adversarial process has been successfully adapted to generate audio data for tasks like sound synthesis~\cite{donahueadversarial}. A prominent application is speech generation from melspectrograms, where models like HiFi-GAN~\cite{kong2020hifi} achieve high-quality speech generation by reconstructing waveforms with exceptional fidelity. GAN has also been applied to generating room impulse responses,~\cite{ratnarajah21_interspeech, fernandez2023generative}. GANs have also shown effectiveness in the recent speech dialogue foundation model~\cite{defossez2024moshi}. 

Anomaly detection leverages GANs to model the data distribution and identify out-of-distribution samples in the test set. Common approaches involve incorporating the learning of inverse mapping in contrast to the generators, which map data back to its latent representation, and using the reconstruction error as an anomaly score. Applications of GAN-based anomaly detection in acoustics include anomalous machine sound detection~\cite{tagawa2021acoustic, jiang2023unsupervised} and deepfake audio detection~\cite{song2024anomaly}.

Additionally, adversarial training has been extended to tasks beyond synthesis, where the generator and discriminator adopt roles tailored to specific acoustic challenges. For instance, the discriminator can be trained as a classifier, while the generator acts as an encoder to extract latent features. In scenarios like channel-agnostic speaker embedding extraction,~\cite{chen2020channel} the discriminator classifies channel information while the generator learns speaker representations. By adversarially maximizing the discriminator’s classification error, the generator successfully encodes features invariant to channel variations, outperforming traditional data augmentation techniques. Similar ideas have been applied to speaker-invariant emotion recognition~\cite{li2020speaker} and audio anti-spoofing~\cite{zhang21ea_interspeech}.

GANs have also inspired novel designs for specific acoustic tasks, such as the design of acoustics metamaterials~\cite{gurbuz2021generative} and underwater noise modeling~\cite{zhou2021generative}. MetricGAN~\cite{fu21_interspeech} focuses on optimizing perceptual metrics for speech enhancement, while other GAN variants tackle super-resolution in audio reconstruction~\cite{eskimez2019adversarial}. These innovations demonstrate how GANs can be tailored and refined for specific applications in acoustics.

\subsubsection{Diffusion Models}

Diffusion models~\cite{sohl2015deep, ho2020denoising} are a class of generative models inspired by thermodynamic processes. They generate samples by a denoising process when they learn to reverse a step-by-step noising process applied to the data. The overall framework, illustrated in the bottom subfigure of Fig.~\ref{fig:generative}, consists of two main stages: a forward diffusion process that adds noise gradually and a reverse denoising process that reconstructs the data. We take a standard diffusion model named denoising diffusion probabilistic models (DDPM) to explain the process in detail. Due to the complexity of the mathematical foundation, we refer interested readers to Ref.~\onlinecite{croitoru2023diffusion} for additional information on other diffusion models and a more generalized formulation through stochastic differential equations. 

In the forward process ($\mathbf{x}_{t-1}$ to $\mathbf{x}_{t}$), Gaussian noise is added to the input data in a series of stages, gradually corrupting it until it is pure random noise. The scale of the noise varies at each step. 
Mathematically, each forward step can be expressed as
\begin{equation}
    q(\mathbf{x}_t | \mathbf{x}_{t-1}) = \mathcal{N}(\mathbf{x}_t; \sqrt{1 - \beta_t} \mathbf{x}_{t-1}, \beta_t I),
\end{equation}
where \( \mathbf{x}_t \) represents the noisy data at step \( t \), and \( \beta_t \) controls the amount of noise added at each step. The coefficients $\sqrt{1 - \beta_t}$ and $\sqrt{\beta_t}$ encourage the distribution of the $t$ step to be closer to a unit distribution compared to the $t-1$ step. The forward diffusion process can be viewed as equivalent to the encoding step of VAE models, which uses latent variables $\mathbf{z}_t$ to estimate the probability distribution.
\begin{equation}
    \mathbf{z}_t = \sqrt{1 - \beta_t} \mathbf{z}_{t-1} + \sqrt{\beta_t} \epsilon_t, 
\end{equation}
where $\epsilon_t \sim \mathcal{N}(0, I)$ is a Gaussian noise.

The reverse procedure ($\mathbf{x}_{t}$ back to $\mathbf{x}_{t-1}$) seeks to restore the original data from the corrupted version by gradually denoising. This is achieved through a neural network that learns to predict and reverse the noise step by step. The denoiser aims to minimize the difference between the predicted clean data and the true data across each step.
\begin{equation}
   p_{\theta}(\mathbf{x}_{t-1} | \mathbf{x}_t) = \mathcal{N}(\mathbf{x}_{t-1}; \mu_{\theta}(\mathbf{x}_t, t), \sigma_{\theta}(t) I)
\end{equation}
where \( \mu_{\theta}(\mathbf{x}_t, t) \) is the model's prediction for the clean data at step \( t-1 \) based on the noisy input at step \( t \), and \( \sigma_{\theta}(t) \) controls the noise removal process.

Latent diffusion models (LDM)~\cite{rombach2022high} have been proposed as a more efficient approach. LDMs learn the denoising process in a lower-dimensional space, rather than directly on the raw data, allowing many irrelevant details in the data to be abstracted away. Consistency models~\cite{song2023consistency} enforce consistency in the generated samples at any step $t$, reducing the number of required steps during sampling and leading to more efficient generation while maintaining high-quality outputs.

In acoustics, diffusion models have been applied in sound field synthesis~\cite{miotello2024reconstruction}, text-to-audio generation~\cite{liu2023audioldm, bai24b_interspeech}, and spatial audio generation~\cite{heydari2025immerse}.

\subsubsection{Discussions on Generative Models}

While VAEs and GANs utilize the same underlying principles, there are some important differences to note.
VAEs stand out with their ability to model complex data distributions with a continuous latent space, offering both high-quality generation and interpretable representations. Their architecture includes encoder and decoder neural networks trained in tandem to enhance representation learning and reconstruction accuracy. GANs are well-known for producing high-quality outputs, where they can create realistic samples from random noise. However, they often require a delicate balance during training, and issues like mode collapse~\cite{srivastava2017veegan} can occur.

In contrast to VAEs and GANs, diffusion models adopt a distinct approach centered around optimizing the reverse diffusion process. This methodology emphasizes learning to predict and remove noise effectively, enabling the generation of high-quality samples that closely resemble the training data without the issues commonly faced by GANs, such as mode collapse or instabilities during adversarial training. Due to their stable training dynamics and flexibility, strong theoretical foundation, and ability to handle complex data distributions, diffusion models are valuable tools for generating realistic data in a variety of domains. The drawbacks of diffusion models include the high dimensionality of the noise (latent variables), which is the same as the original data, and the slow inference speed resulting from the large number of steps involved in the sampling process.

\subsection{Implicit Neural Representation}

Traditional data representations often take the form of high-dimensional matrices. For example, an image is typically stored as a matrix of pixel values, where the dimensions correspond to its width and height. However, this representation poses limitations, particularly when merging datasets of the same category but with varying resolutions and dimensions. Such structural inconsistencies can complicate downstream processing and integration tasks and hinder the generalizability of models trained on data with fixed dimensions.

Implicit neural representations (INRs),~\cite{sitzmann2020implicit} also referred to as neural fields, offer a continuous and differentiable method for representing discrete data using neural networks. Instead of explicitly storing data in a matrix, INR uses a neural network to map input coordinates to corresponding output values, creating a continuous data representation. For example, in the case of images, a small neural network is trained to represent a single image. The network takes spatial coordinates as input and outputs the corresponding RGB pixel values at those coordinates. This approach inherently supports interpolation and enables the seamless merging of datasets with different resolutions.

The mathematical formulation is 
\begin{equation}
      f_\theta(\mathbf{x}) = \mathbf{y},
\end{equation}
where
$ f_\theta $ is the neural network parameterized by \(\theta\),
$ \mathbf{x}$ is the input coordinate (e.g., spatial coordinates),
$ \mathbf{y}$  is the output value (e.g., amplitude, color, distance, or any other property).

This representation can potentially model the distribution of a specific data type across varying dimensions or resolutions. It also relates directly to meta-learning, enabling generalization across multiple INRs. This approach contrasts with traditional neural network methods, which are typically trained on large datasets with fixed input-output structures. In the conventional case, the neural network is trained to produce an entire output of fixed dimensionality given some conditional input.

INRs have primarily evolved within computer graphics and computer vision over the years, but they have recently been adopted in acoustics, particularly in spatial audio, due to the location-dependent nature of the data and its requirement for consecutive resolution.

Creating immersive audio experiences requires accurate modeling of sound propagation from the source to the listener through space. A key challenge in room acoustics is modeling room impulse responses. However, real-world measurements are limited by the microphone-loudspeaker setup, which can only capture a restricted number of source-listener pairs. Traditional methods use mathematical interpolation to calculate the impulse response at new locations, while INRs provide a novel solution. Neural Acoustic Fields (NAFs)~\cite{luo2022learning} were proposed to leverage neural fields to represent impulse responses in the time-frequency domain. Implicit Neural Representation for Audio Scenes (INRAS)~\cite{su2022inras} further refined this approach, introducing a decoupled module model to represent the scatter-bounce-gather process in audio propagation. This research has recently been extended to audio-visual novel-view acoustic synthesis,~\cite{liang2024av} where camera angle information from visual cues is incorporated as input to predict the impulse response at specific audio-visual scenes, making it particularly useful for audio-visual navigation.

The binaural audio through headphones or VR headsets is rendered not only with room acoustics but also incorporates the propagation to the ears, described by head-related transfer functions (HRTFs), and faces a similar modeling challenge due to limited measurements.  HRTFs are continuous functions that take spatial directions as input and output the spectrum across all frequency bins and thus INRs are well-suited for this task. This idea was applied to binaural audio,\cite{gebru2021implicit} where personalized HRTFs were implicitly modeled by estimating transformation functions for binaural synthesis using neural networks. The measurement directions do not constrain this method and directly predict binaural audio, with HRTFs as intermediate outputs and no ground truth. 
HRTF Field~\cite{Zhang2023HRTFfield} directly applies INRs to model HRTFs across datasets and reveals another benefit of mixed database training for interpolation tasks, alleviating differences in spatial sampling schemes. This approach was further extended~\cite{Yoshiki2024niirf}, where the model estimates the coefficients of cascaded infinite impulse response (IIR) filters rather than the HRTF magnitude directly, enabling a more compact representation that better captures the resonant characteristics of HRTFs with fewer parameters.

INRs are memory-efficient due to their simple architecture and ability to represent infinite resolution with the same set of parameter weights. However, optimizing INRs is challenging because they rely on continuous representations. INRs are prone to overfitting if training data is insufficient or lack diversity, which requires proper regularization to generalize well, such as mitigating differences between the HRTF databases~\cite{wen2023mitigating}.  INRs may struggle to accurately capture highly detailed or sharp features, particularly in data with high-frequency content.

\subsection{Physics-informed Machine Learning}
Physics-informed machine learning (PIML) integrates physical principles with ML to solve scientific and engineering problems~\cite{karniadakis2021physics}. Many physical systems are governed by physical laws that are (partially) understood through centuries of scientific progress. It is only natural to leverage well-established scientific knowledge to improve current ML workflows. At the same time, ML can be used for scientific discovery~\cite{rudy2017data,li2023data}, helping us understand physical aspects that are currently poorly understood or too complex to model with traditional methods.

Incorporating physical knowledge into ML models can improve their accuracy, efficiency, interpretability, robustness, and generalization capabilities. Physical priors guide models toward learning physically plausible solutions, making them more accurate than ML models that rely purely on data. By adding physical constraints, the space of possible models considered by the learning algorithm is narrowed. Consequently, PIML models tend to be data-efficient, making them particularly useful in scenarios where data is scarce or expensive. In contrast, traditional ML models typically require large amounts of training data.

Physically motivated constraints also act as effective regularizers, improving model robustness. Models that incorporate physical laws generalize better and can extrapolate to regions where data is sparse or unavailable. On the other hand, traditional ML models often perform poorly outside the range of the training data and are more prone to overfitting.

The interpretability of ML models---i.e., the ability to understand and explain how the models make predictions and decisions---is crucial for building trustworthy ML systems. PIML models are generally more interpretable because they adhere to physical laws, leading to more reliable predictions and a better understanding of the model's behavior. In contrast, traditional ML, especially deep learning models, are often considered `black boxes' with limited interpretability. Additionally, PIML models often require fewer parameters and less complex architectures compared to traditional ML models. These simpler models are often more transparent and easier to interpret.

There are different strategies to incorporate physics into ML workflows. One way is to embed physical principles directly into the network architecture design. An example of this is neural ordinary differential equations~\cite{chen2018neural}, 
which link residual neural networks to numerical time integrators. Following this idea, a very active field of research involves the design of custom network architectures that can predict the time evolution of dynamical systems robustly and efficiently~\cite{zhai2023parameter}.
Another way of incorporating physics into ML is to include physical constraints in the loss function, as described in the following section. 

\subsection{Physics-informed neural networks}
\begin{figure}
    \centering
    \includegraphics[width=1\linewidth]{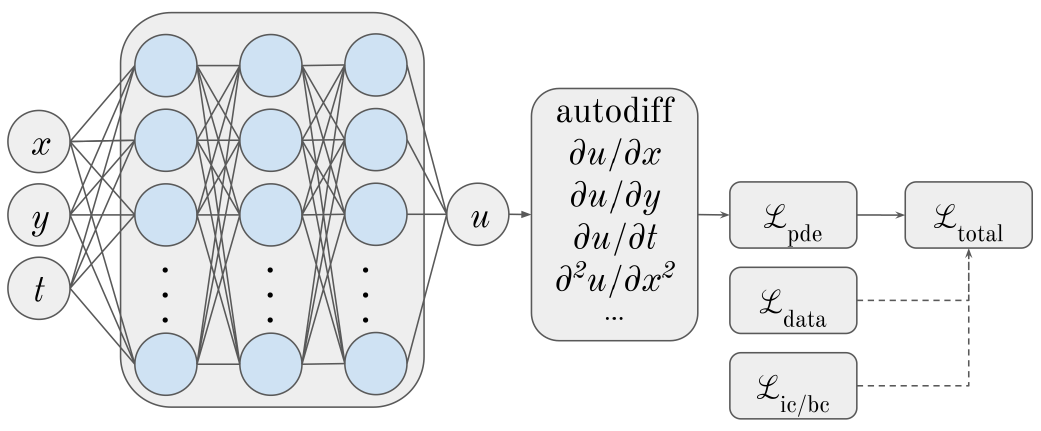}
    \caption{Diagram of a PINN. The neural network inputs are the coordinates $x,y,t$, and the output is the physical quantity of interest $u(x,y,t)$. Automatic differentiation is used to compute the partial derivatives of the output with respect to the inputs. A physics loss term, $\mathcal{L}_\text{pde}$, that contains the underlying PDE is formed, and added to loss terms that account for initial/boundary conditions, $\mathcal{L}_\text{ic/bc}$, and/or observed data, $\mathcal{L}_\text{data}$, to compute the total loss, $\mathcal{L}_\text{total}$.}
    \label{fig:pinn}
\end{figure}

This section provides an introduction to \textit{physics-informed neural networks} (PINNs), one of the most popular modes of PIML. PINNs are neural networks that integrate physical constraints into their loss function. Physical systems are often expressed as partial differential equations (PDEs). These can be linear, such as the acoustic wave or Helmholtz equations, nonlinear, such as the Burgers' equation, or a system of coupled PDEs. PINNs approximate the solution of a PDE by incorporating a residual term into its loss function that contains the PDE. Fig.~\ref{fig:pinn} shows an example of a PINN. During training, the PDE residual is minimized, along with other terms that account for initial/boundary conditions and observed data. A key aspect of PINNs is the use of automatic differentiation to compute the differential equations that encode the physics into the loss function. Automatic differentiation, the backbone of modern ML, makes it possible to easily compute partial derivatives by breaking functions into elementary operations and applying the chain rule systematically.

The term PINNs was popularized around 2019 ~\cite{raissi2019physics}. Since then, PINNs have been applied in many fields of science and engineering~\cite{karniadakis2021physics,cuomo2022scientific}, including fluid dynamics~\cite{cai2021physics}, climate modeling~\cite{kashinath2021physics}, and biomedicine~\cite{kissas2020machine}, to name a few. In acoustics, PINNs have been applied in ocean acoustics~\cite{yoon2024predicting}, atmospheric sound propagation~\cite{pettit2020physics}, room acoustics~\cite{borrel2021physics,karakonstantis2024room}, spatial audio and sound field control~\cite{koyama2025physics}, acoustic holography~\cite{olivieri2021physics},  ultrasound imaging~\cite{shukla2020physics,wang2023acoustic}, nonlinear propagation~\cite{savovic2023comparative}, and spatial inverse problems~\cite{liu2024spatial}.

As with other PIML approaches, PINNs achieve better generalization and require less data than purely data-driven neural networks while being expressive enough to approximate complex PDE solutions. Unlike conventional numerical methods, PINNs are gridless, i.e., they can make predictions at any resolution without the need to be retrained. Since the PDE is enforced over the full domain, PINNs can make zero-shot predictions, i.e., the solution can be predicted at points where there are no data. 
In addition, PINNs are highly flexible, allowing them to solve both forward and inverse problems. For instance, a forward problem involves computing a wavefield given initial and boundary conditions, while an inverse problem aims to estimate PDE parameters (e.g., wave speed profile) from observed data. AcousticsML includes notebooks for forward and inverse problems involving the wave equation.

However, PINNs have limitations and challenges. Training a PINN for solving a forward problem is significantly slower than using conventional numerical solvers such as finite differences or finite elements. This is due to the need to extend the computational graph to compute the partial derivatives that constitute the PDE residual. Further, training PINNs can be difficult due to competing terms in the loss function and gradient stiffness ~\cite{wang2021understanding,wang2022and,rathore2024challenges}. Moreover, like other deep neural networks, PINNs suffer from spectral bias, struggling to capture the high-frequency content of the PDE solution~\cite{wang2021eigenvector}.

Several extensions have been proposed to address these challenges, making PINNs a very active field of research. Actively selecting training points can achieve faster convergence~\cite{nabian2021efficient}. Annealing algorithms that automatically scale different terms in the loss function have been proposed to alleviate gradient stiffness issues~\cite{wang2021understanding}. The use of Fourier features~\cite{wang2021eigenvector} and subdomain partitioning~\cite{jagtap2020extended,moseley2023finite} has been proposed to address spectral bias and multiscale problems. Some of these extensions are covered in the AcousticsML notebooks. 

\subsection{Hyperparameter optimization}

Most ML algorithms are controlled by tunable parameters that the user sets.
Such parameters are often referred to as hyperparameters, as they are distinct from the parameters---or model weights---that are learned during the training process.
For example, neural network weights are learned during training, while the learning rate, number of layers, number of neurons per layer, and other settings are hyperparameters set by the user.
Other examples of hyperparameters include the number of trees in a random forest, the number of clusters in a clustering algorithm, and the number of neighbors in a nearest neighbors algorithm.
More generally, hyperparameters are any parameters not learned during the training process and must be set by the user.
This can include the choice of algorithm, loss function, or preprocessing steps, although in practice, such design choices are typically informed by domain knowledge.

Hyperparameter optimization is critical to selecting the model that best explains the dataset according to some criterion.
The choice of criterion typically includes some measure of model performance, such as accuracy, precision, recall, or error, and may also incorporate computational cost.
Consider a ML model $F_\theta$ with trainable parameters $\theta$ that maps data from an input space $\mathcal{X}$ to predictions in an output space $\mathcal{Y}$:
\begin{equation}
    F_\theta : \mathcal{X} \rightarrow \mathcal{Y}.
    \label{eq:mapping}
\end{equation}
The objective of hyperparameter optimization is to find the best set of hyperparameters $\phi^*$ that minimizes a loss function $\mathcal{L}$ over the hyperparameter space:
\begin{equation}
    \phi^* = \arg\min_\phi \left[\mathcal{L}(F_\theta)\right]_\phi.
    \label{eq:optimal_hyperparameters}
\end{equation}
Eq.~\eqref{eq:optimal_hyperparameters} can be solved in several systematic ways.
A common approach is grid search, in which each hyperparameter is assigned a set of possible values, and the model is trained and evaluated for each combination of hyperparameters.
This approach is simple and intuitive, but can be computationally expensive or even prohibitive, especially for models with many hyperparameters.
Random or quasi-random searches can also be used, but may still require many models to be trained and evaluated\cite{bergstra2012random}.
Recent advances in Bayesian optimization have shown promise in addressing these challenges by constructing a probabilistic surrogate model of Eq.~\eqref{eq:optimal_hyperparameters} and using the surrogate model to select the next set of hyperparameters to evaluate.
Bayesian optimization incorporates the results of previous evaluations to inform the selection of future hyperparameters, allowing for more efficient exploration of the hyperparameter space\cite{shahriari2016taking,jenkins2024geoacoustic}.
Grid and random search are readily implemented in many ML libraries, while Bayesian optimization is available in specialized ML frameworks\cite{martinez-cantin2014bayesopt,akiba2019optuna,balandat2020botorch}.

\subsection{Uncertainty quantification}

In physical systems, obtaining the parameter uncertainty, i.e., the degree to which the parameters are unknown, is nearly as important as obtaining the parameter estimates. Yet, most ML in acoustics neglects this.
ML is grounded in the training-testing paradigm, in which the model parameters are estimated to minimize a loss function and then validated with test data. There is a consensus that ML models are more accurate than simple statistical models when making predictions due to their flexible nature, as nicely advocated in ``Statistical modeling: the two cultures'' \cite{breiman2001statistical}.  The advent of new loss functions tailored to estimate probability distributions, combined with the progress in ML, leads to ML models that can estimate predictive uncertainty more accurately than simpler statistical models.

Uncertainty can be reducible (epistemic or statistical uncertainty) and irreducible (aleatoric or systematic uncertainty). The reducible uncertainty can often be reduced by, e.g., collecting more data and averaging, while the irreducible uncertainty can be mitigated by replacing the data model with a more robust alternative.
Both uncertainties can be reduced by careful design, and an indication of whether this is successful can be obtained by analyzing the resulting uncertainty.

Uncertainty quantification (UQ) involves identifying sources of uncertainty, assessing their impact on model outputs, and providing a measure of confidence in the model's predictions. UQ methods can generally be divided into Bayesian and frequentist methods. 

Bayesian methods use a prior distribution, which describes our prior knowledge about the parameters, and a likelihood distribution, which describes the probability of the observed data given the parameters, to obtain a posterior distribution via Bayes' theorem\cite{gerstoft1998ocean,bonnel2013bayesian}. The uncertainty intervals are then obtained from the posterior distribution, called credible intervals (region of the posterior distribution containing $1-\alpha$ of the probability). %\cite{huang2006validation}. 
Credible intervals are used in Bayesian statistics to characterize the uncertainty of an unobserved parameter. For example, a 95\% credible interval means there is 95\% probability that the parameter lies within this range, given the data and the prior information. 

Frequentists assume that the true unknown estimate is fixed and the uncertainty is quantified in terms of confidence intervals. Confidence intervals are derived only from sampled data, without a prior distribution. For a chosen confidence level $1-\alpha$, after running $N$ tests with $N$ confidence intervals, $1-\alpha$ of the confidence intervals are expected to include the true value.
For example, a 95\% confidence interval means that if the experiment were repeated many times, 95\% of the intervals from those experiments would contain the true value. Methods for calculating confidence intervals include bootstrapping (repeated resampling)\cite{tibshirani1993introduction} or direct interval estimation by assuming an output distribution. % \cite{lu2012analysis}.

The credible intervals obtained using Bayesian methods represent the level of uncertainty associated with a random variable.
Bayesian methods make assumptions for the prior distribution, which might be restrictive. This and non-linear forward models make a closed-form expression for the posterior distribution difficult to obtain. Sampling techniques can approximate the posterior distribution but lead to computational overhead.

Bayesian sampling has a rich history in acoustics \cite{gerstoft1998ocean,dosso2002quantifying,bonnel2013bayesian,xiang2020model,vardi2024estimation}. These can provide an accurate sampling of the probability distributions, though often with some bias due to the choice of sampling parameters. They are computationally demanding, as accurate sampling requires many forward model runs. A simpler strategy is to perform UQ, providing a measure of uncertainty for the parameter estimate or observations.
Although many UQ methods are available, we focus on interval estimation methods through the recently introduced prediction intervals with conformal prediction (CP)\cite{khurjekar2023uncertainty,khurjekar2024multi}.

\begin{figure}
\centering
\includegraphics[width=1\linewidth]{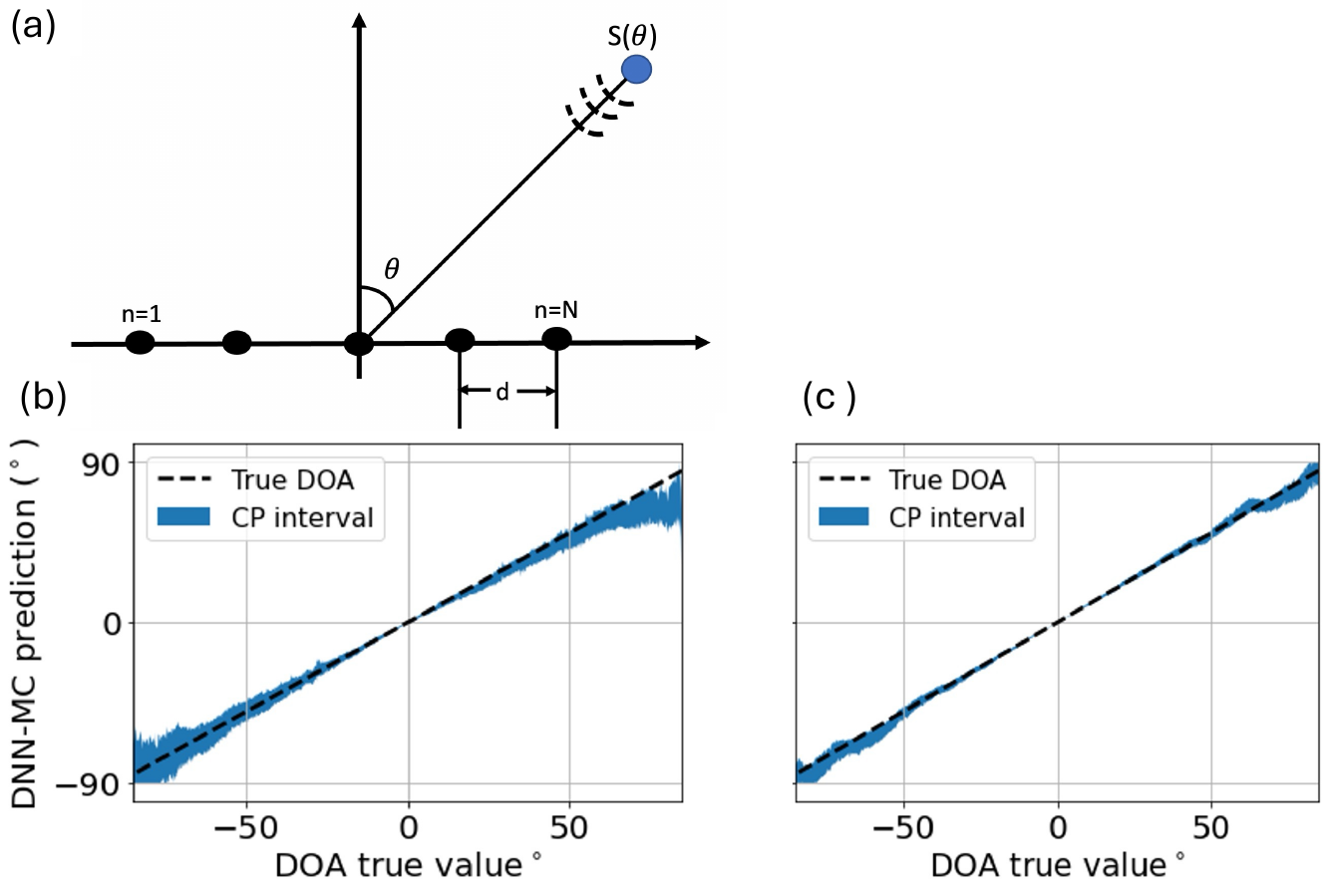}
\caption{Beamforming example of prediction interval with (a) setup. Conformal Prediction interval versus true direction of arrival (DOA) for SNR of (b) 0 dB and (c) 20 dB.
\label{fig:CP}}
\end{figure}

CP computes the prediction intervals in a few steps \cite{shafer2008tutorial,angelopoulos2021gentle} as indicated in the example in Fig.\ \ref{fig:CP}.
CP utilizes a parameter estimate plus a heuristic measure of uncertainty (scalar) to define a conformal mapping between this uncertainty scalar and the end points of a prediction interval that contains the true estimate with probability $1-\alpha$ based on {\it just one observation}. 
The conformal mapping refers to an unknown angle-preserving nonlinear mapping between these quantities, as the mapping is unknown and thus has to be learned via training data.
 
We demonstrate the approach with a simple example \cite{khurjekar2024multi}. Consider estimating the direction of arrival (DOA) $y$ from observations on an array of observations $\bf x$,
\begin{equation}
y=f({\bf x}),
\end{equation}
where the $f$ could be any beamformer estimating a DOA, such as conventional beamforming or the DOA output of a neural network. We first train a neural network to give estimates of the DOAs.
For a single observation $i$ with input ${\bm x}_{i}$ received, a neural network estimates the mean DOA $\mu_i$ and the estimated variance $\sigma^2_{i}$ obtained by running the trained neural network with dropout, where the dropout is used for estimating a heuristic variance $\sigma^2_{i}$. This uncertainty estimate $\sigma^2_{i}$ is not guaranteed to satisfy the statistical coverage. CP can remedy this issue by calibrating the estimate using training data and conformal mapping.
  
The prediction interval for a single test point $\textbf{x}_{i}$ with estimated $\mu_i$ and variance $\sigma^2_{i}$ is, 
 \begin{equation}\label{eq:CPint}
    \mathcal{C}(\textbf{x}_{i}, \alpha) = [\mu_i^k - \sigma_i {q}_{\alpha},~\mu_i^k + \sigma_i {q}_{\alpha}].
\end{equation}
Each DOA direction obtains the calibration factor ${q}_{\alpha}$. To obtain ${q}_{\alpha}^k$, we generate $L$ realizations of ${\bm x}$ for random DOA $y^{\rm true } \in [-90^\circ, 90^\circ]$ with random noise added at a given signal to noise ratio and estimate the $\mu^l$ and  $(\sigma^2_{})^l$.

Defining a score function as $| \mu^l-y^{\rm true } |/ \sigma^l$,
then we rank the $L$ scores and pick the $1-\alpha$ ratio so that
\begin{equation}
{\rm Prob}\left[ \frac{| \mu^l-y^{\rm true } |}{\sigma^l} \le {q}_{\alpha}\right]\ge 1-\alpha.
\end{equation}
This determines the ${q}_{\alpha}$ for the whole dataset.

Figure \ref{fig:CP} demonstrates the CP prediction on a simple DOA estimation for a 20-element linear array with half-wavelength spacing for one source.
The data $\bm x$ are generated knowing the true DOA and adding noise to this sample. 
%The confidence and credibility intervals would give a constant error estimate, symbolized by the squared error estimate.
CP gives an uncertainty interval for just one observation, as seen for varying across the observed direction of arrival, see Fig.\ \ref{fig:CP}(b) and (c). The uncertainty interval increases for lower SNR ratio, see (b) vs.\ (c).

\subsection{Explainable AI}
\begin{figure}
    \centering
    \includegraphics[width=0.8\linewidth]{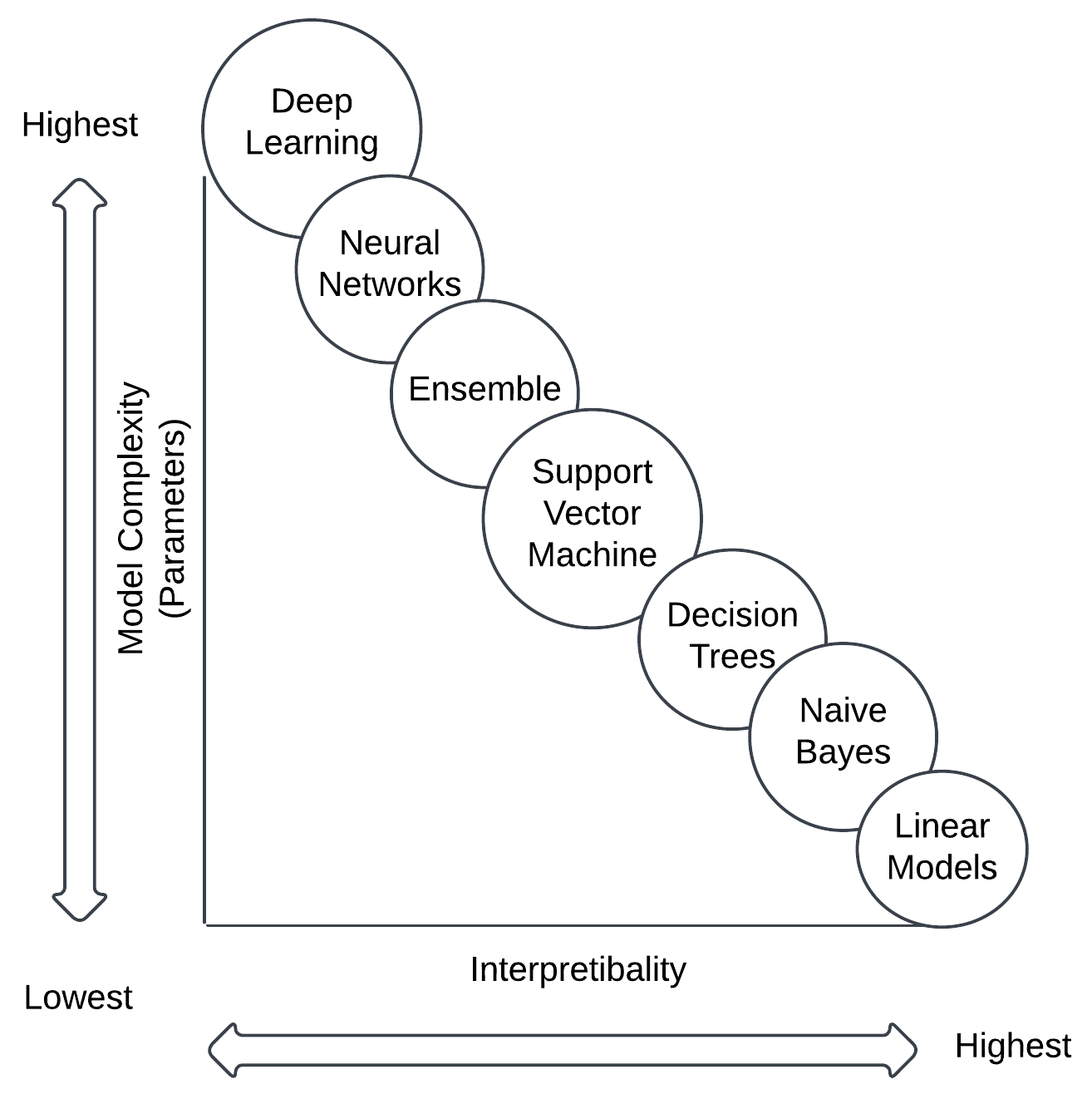}
    \caption{Model complexity vs.\ model interpretability. The larger the number of neurons, nodes, or higher-order fits, the more difficult it is to interpret the results. }
    \label{modelinterpret}
\end{figure}

Larger and more complex models have recently become popular for many applications due to their higher accuracy than traditional ML models, automatic feature engineering, and ability to learn complex features. Understanding how models make predictions becomes crucial for building trust and enabling human oversight as these models grow in complexity. Models are often described as ``black box'' diagrams because of their complexity and lack of transparency in their predictions. Examples of the model complexity are shown in Fig.~\ref{modelinterpret}, where models are plotted as a function of complexity (i.e., number of hyperparameters) vs.\ the level of interpretability. Interpreting how these ML models achieve their success can be non-trivial. Statistical performance measures such as accuracy do not provide enough key information for model interpretation or reliability. Recently, AI has sparked increasing research interest in explainability and explainable AI \cite{dwivedi2023explainable,angelov2021explainable}, which can be broken down into a few different forms: 1) data explainability, 2) model explainability, 3) feature-based explainability, and 4) example-based explainability. We briefly discuss these varieties and a few explainable AI techniques.

\subsubsection{Data Explainability}

Data explainability focuses on the data used to train and input into a model. This includes identifying biases and underrepresented samples. One approach is data visualization to identify patterns, trends, and insights. Visualizing the interconnectivity between desired outputs and inputs is difficult for larger datasets thus a popular approach is to use dimensionality reduction techniques such as principal component analysis (PCA), t-distributed stochastic neighbors (t-SNE), uniform manifold approximation and projection (UMAP) \cite{mcinnes2018umap}, TriMAP \cite{amid2019trimap}, dictionary learning, autoencoders, or pairwise controlled manifold approximation (PacMAP)\cite{wang2021understandingpacmap}. Algorithms provide unique dimensionality reduction techniques focusing on finding relations in the data mapped to a new dimensionality that reduces the variance. It should be cautioned that they may not represent the higher dimensionality correctly and could result in nonexistent relations based on the number of newly mapped dimensions. Further details can be found in Chapter 6 of the AcousticsML repository.

\subsubsection{Model Explainability}

An ML model's capacity and complexity are due to the number of learnable parameters. As the number of parameters increases, it becomes difficult to interpret a model's prediction (see Fig.~\ref{modelinterpret}). Although larger models can produce highly accurate results, understanding model predictions can be ambiguous, thus referring to these models as ``black box" models. Examples of black box models include deep neural networks, convolutional neural networks (CNN), and large gradient boosting models. Alternatively, smaller, more transparent, and interpretable models for prediction are referred to as ``white box" models. White box models include linear models, Gaussian mixture models (GMM), Naive Bayes models, and decision trees. These models offer rule-based decisions and simple equations learned from training data. The trade-off for higher interpretability and transparency is prediction accuracy in these models. The choice of model complexity depends on the application and the desired outcome. 

\subsubsection{Feature-Based Explainability}

Features are measurable properties that are input to a system. These inputs are related to a particular data sample that an ML model can interpret to make predictions. Features are typically independent and provide specific details about the sample. In acoustics, features can be 1D representations (e.g. intensity, energy, timbre, etc.) or 2D representations (e.g. spectrograms) for a given sample. Descriptions of features and feature selection are in Chapter 2 of the AcousticsML repository.

Feature selection can reduce complexity and improve ML model accuracy. Providing vague or correlated information can lead to misleading results, such as passing vague information about energy in frequency bands for classification. To improve model accuracy, we can consider the relative importance of each input for the given outputs, known as feature importance. One approach is to use prior statistical tests (e.g., Chi-squared test), or correlation tests to eliminate features before training a model to reduce the complexity. These approaches can help identify features with little variability that do not contribute significantly to the output. Feature importance can be learned after training a model using random permutations, feature nulling, or Recursive Feature Elimination. These techniques compare the accuracy of the model predictions before and after altering the inputs. Similarly, feature weights (i.e., coefficients for linear regression or number of occurrences for decision trees) may be beneficial in determining feature importance. For further information on feature selection approaches, see Ref.~\onlinecite{stanczyk2014feature}.

\subsubsection{Example-Based Explainability}

Example-based explainability aims to identify how models make predictions by looking at global or local predictions. Global predictions provide a broad analysis of inputs, while local predictions focus on how smaller sample sets are predicted from the given inputs. Global techniques for identifying how models make predictions include random permutations \cite{fisher2019all}, accumulated local effects \cite{apley2020visualizing}, or partial dependence-based feature importance [~\onlinecite[Sec.~18.6.2]{murphy2022probabilistic}]. Local techniques include Local Interpretable Model-Agnostic Explanations (LIME)\cite{ribeiro2016should}, Shapley Additive explanations (SHAP)\cite{lundberg2017unified}, or anchors \cite{ribeiro2018anchors}. Some of these techniques are covered in the tutorial notebooks for unsupervised, supervised, and DL models in Chapter 6 of the AcousticsML repository.

Despite its growing potential, explainable AI has several limitations and challenges. One of the most fundamental issues is the trade-off between model accuracy and interpretability: complex models as DNNs, which outperform simpler ones but operate as "black boxes," offer little insight into how decisions are made. Although \emph{post hoc} explanation techniques such as SHAP or LIME attempt to provide interpretability, these methods can produce inconsistent or misleading interpretations that do not reflect the internal logic of the model. The interpretability of model predictions often depends on the specific application, and the choice of explanation method is typically guided by the user's particular needs and objectives. Another key challenge is integrating domain-specific knowledge to enhance both model performance and interpretability without introducing bias or overfitting. The current lack of standard evaluation metrics for explanations complicates efforts to compare or validate explainable AI methods. In certain acoustic applications, this becomes particularly problematic, as flawed or opaque explanations can undermine trust, impede accountability, and lead to invalid predictions. The evolving nature of AI necessitates explanations that are adaptive and context-aware, capable of keeping pace with changing models, data distributions, and evolving ethical standards. 

\section{The AcousticsML Repository}
\begin{figure*}
    \centering
    \includegraphics[width=\linewidth]{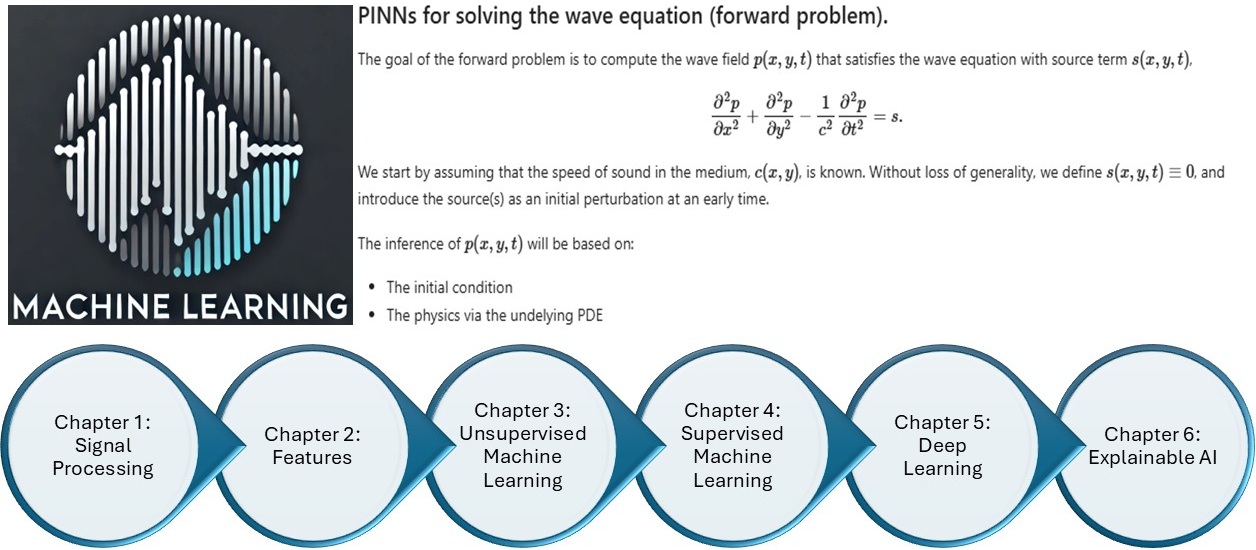}
    \caption{Overview of AcousticsML. AcousticsML includes ML examples for Acoustic Applications and is split into chapters (bottom), which include descriptions of the application and ML model used (top).}
    \label{fig:GitHubOverview}
\end{figure*}

Due to the substantial amount of acoustic applications and ML models, the AcousticsML repository addresses particular topics that can be applied to broader applications. The AcousticsML repository provides an overview of the topics covered at the top of the page, followed by a brief discussion of the model used, how models are initiated and trained, and further references for further information (Fig.~\ref{fig:GitHubOverview}). Notebooks in the AcousticsML repository are grouped into six chapters that introduce ML techniques and applications that can be extended to other applications.

{\bf Chapter 1) } Short introduction to signal processing techniques. Signal processing enables models to learn from data efficiently and effectively. Though the notebook does not provide all information on the theory of waves, ray propagation, or acoustic modes, it gives a brief background, additional links, and learning resources.

{\bf Chapter 2) } Feature extraction and selection from acoustic data. The notebooks briefly discuss 1D statistical measures and 2D spectral features that can be input into a model.

{\bf Chapter 3) } Unsupervised ML algorithm applications. These algorithms learn from the data and do not require prior labels, revealing patterns and relations in the data that may be difficult to recognize.

{\bf Chapter 4) } Supervised ML algorithm applications. These models learn from labeled data to predict desired outcomes. 

{\bf Chapter 5) }Deep learning model applications. Notebooks emphasize DL through PyTorch and demonstrate CNNs, GANs, and PINNs. 

{\bf Chapter 6) } Explainable AI techniques for unsupervised, supervised, and DL models. Explainable AI techniques are significant to today's acoustic applications, providing insights into model prediction and human interpretation of data.

\begin{figure}
    \centering
    \includegraphics[width=0.7\linewidth]{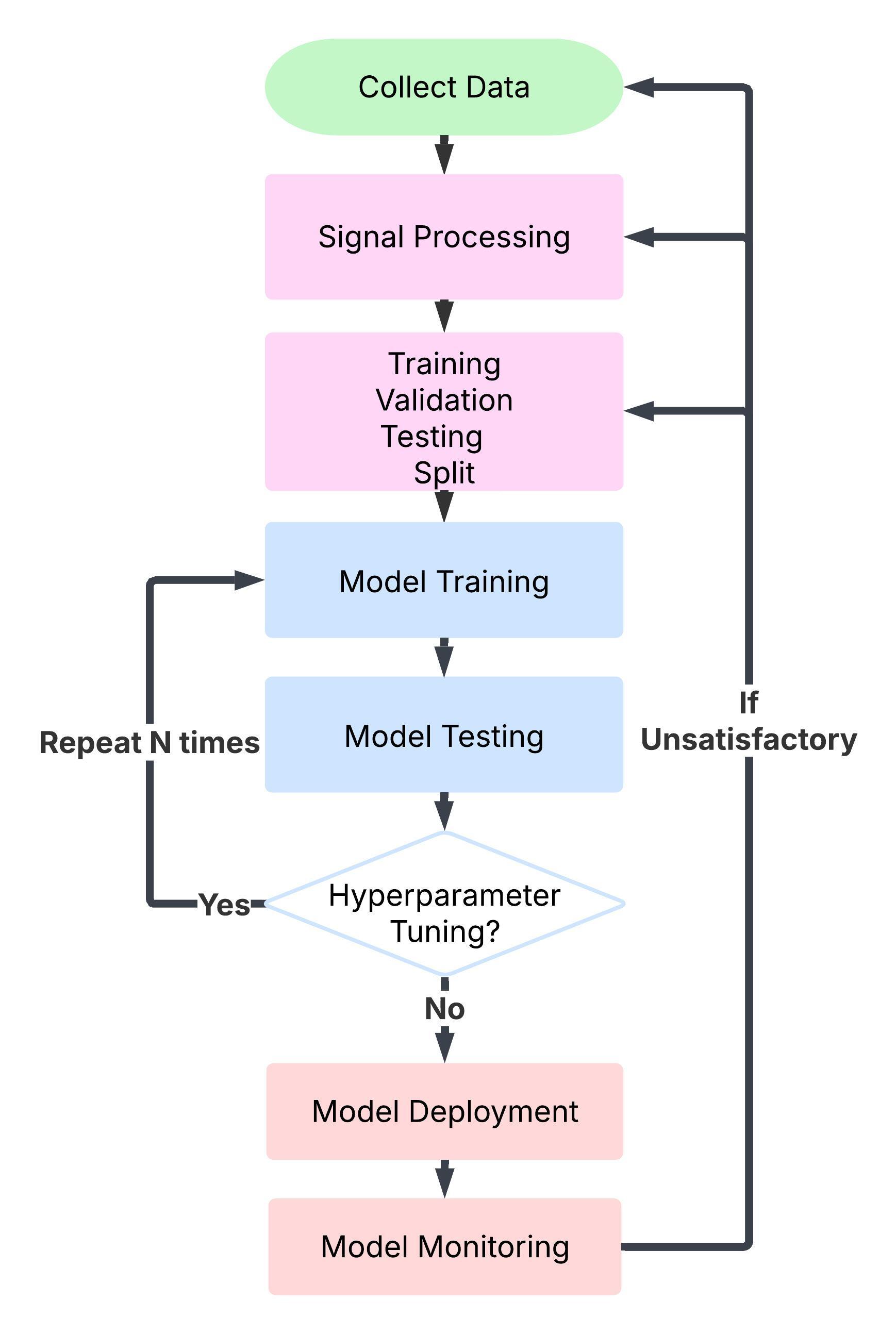}
    \caption{Workflow of ML models. Models are trained through a series of steps and iterated based on a performance metric to improve the model's capability.}
    \label{workflowexample}
\end{figure}

The AcousticsML repository follows a typical ML workflow illustrated in Fig.~\ref{workflowexample}. First, data are selected and preprocessed using signal processing techniques. This process provides quality assurance and quality control (QA/QC) to remove random noise in the data, select relevant data, or improve model performance. The features are then extracted from the data using several quantitative and qualitative techniques to train the ML model algorithms. 

Several ML model architectures are available including unsupervised, supervised, and DL. The choice of model and implementation depends on the application, the amount of labeled data available, and the desired implementation for prediction (e.g., regression or classification). Notebooks do not cover which model architecture to use but provide examples of some available models to get started. Trained models are tested with additional available data to ensure they operate efficiently and successfully. This step is vital in determining whether a model was trained effectively and is generalizable. Training and testing models can occur repeatedly, adjusting hyperparameters to test performance. An additional post-processing step can be included to improve model prediction, but this is left out in this example. 

Once a trained model has satisfactory performance, it can be deployed. Deployed models can be monitored with newly collected data to observe biases and prediction errors. If performance deteriorates over time, it may be due to noisy data, new unseen observations (i.e., observations not present in the training dataset), or the limited ability of the model. In any case, training a new model with the newly collected data, different signal processing techniques, or a new model architecture may be advantageous. As an important note, there is no one way to approach a problem with ML; instead, sets of techniques and model architectures can be applied effectively.

\section{Application Example Workflows}\label{appexampworkflow}
We highlight several acoustic examples demonstrating ML pipelines. Pipelines describe the procedure of preprocessing, data input, and output from an ML model, as shown in Fig.~\ref{workflowexample}. The AcousticsML repository examples provide initial workflows and procedures to learn and apply to other applications. Four applications and unique machine-learning models are selected to discuss their procedures and applications. The code for each application is available as a Jupyter Notebook in the AcousticsML repository.

\subsection{Acoustic Classification}

In acoustic classification tasks, acoustic data are assigned to predefined or learned classes according to the features of each data sample. Classification can be applied to distinguish different kinds of animal calls \cite{BirdNET, Kahl2017}, identify musical instruments \cite{blaszke2022musical}, classify environmental sounds (e.g., anthropogenic noises or bioacoustics) \cite{Bermant2019}, or monitor for malfunctions in machinery \cite{tama2023recent,li2020intelligent,saufi2019challenges}.
Classification leverages existing datasets to predict or classify new or unseen datasets into distinct classes based on similarity or probability. Deep learning has become a powerful tool for sound classification, enabling models to automatically learn complex features from raw audio data without manual feature engineering. CNNs, recurrent neural networks (RNNs), and, more recently, transformers are commonly used architectures for this task, leveraging large datasets and identifying non-linear patterns between inputs to make predictions. Choosing which model to implement depends on several factors, including the complexity of the problem, the amount of data available for training, and the choice of features to be used for prediction.

A challenge in acoustic classification is ensuring that training data is available to represent each class adequately. A model cannot classify something on which it is not trained; hence, the more diverse a classification problem (e.g., predicting a bird species from birdsong), the more training examples from each class are required to train a model. When little data is available to train a model, anomaly detection can first be used to identify sounds of interest, and unsupervised or manual labeling can be used to group similar sounds in subsequent analysis.

The choice of features used to classify acoustic data may also impact the performance of a model. For example, models may have difficulty differentiating between two animal species with similar vocalizations (i.e., similar frequency upsweeps, duration, loudness, etc.).
Acoustic classification, therefore, depends on thoughtful feature engineering, where the representation of sound data directly impacts models' discriminative power.
For instance, time-frequency representations like spectrograms may outperform simple frequency spectra by preserving temporal dynamics that reveal distinctive patterns in acoustic events.
Feature engineering requires systematic experimentation to identify which acoustic features or transformations most effectively capture the discriminative characteristics of the target sounds.

Here, we describe an example of an acoustic classification approach and demonstrate how the model choice, amount of data used for training, and feature representation impact performance. Chapter $4$ in the AcousticsML repository includes several Jupyter Notebooks demonstrating the application of classification to labeled acoustic data.

\subsubsection{Dataset}
Audio classification is demonstrated on data from the Audio Modified National Institute of Standards and Technology (AudioMNIST)\cite{becker2024audiomnist}. This dataset consists of 30,000 recordings of spoken numbers in English, with 50 repetitions from 60 speakers of different nationalities. The duration of audio clips ranges from 0.3--1 s, with $3,000$ examples recorded for each spoken number. All audio clips are sampled at $22.05$ kHz, and two recording locations are used. %(shown in Fig.~\ref{fig:audiomnsitlen} top)

\subsubsection{Feature Extraction}
\begin{figure}
    \centering
    \includegraphics[width=\linewidth]{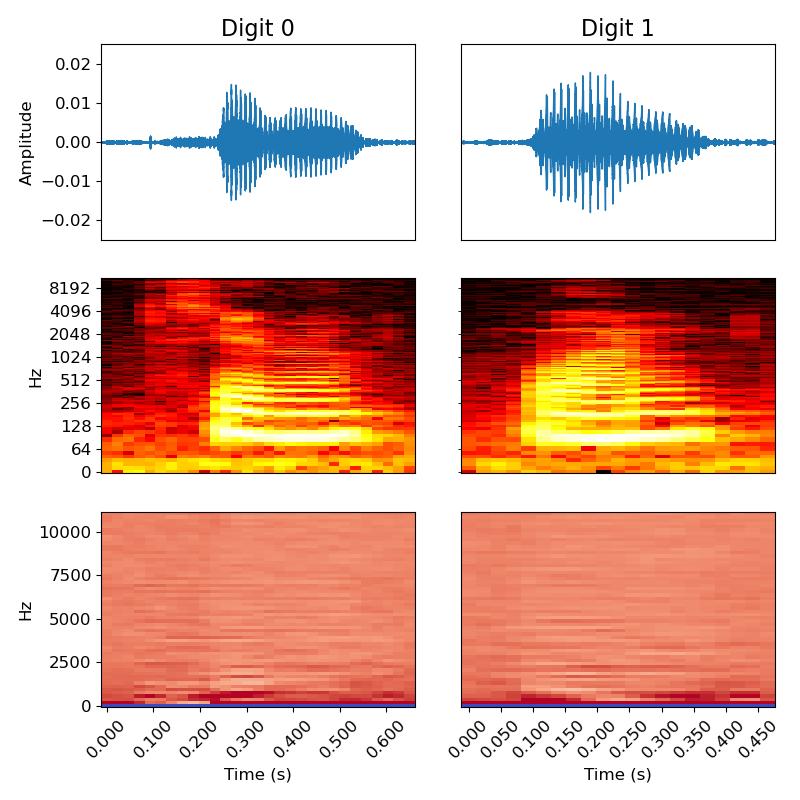}
    \caption{Spectra features from audio examples. Examples of audio data (top row) from the \mbox{AudioMNIST} dataset, with features extracted by FFT (middle row) and Mel-frequency cepstral coefficients (bottom row).}
    \label{fig:example_extraction}
\end{figure}

Two feature extraction techniques were evaluated to identify key differences between each spoken number. Both techniques transform the raw audio data into 1 by 1026 feature vectors. The first technique, a Fast Fourier Transform (FFT) approach, extracts features from frequency components in each audio clip from a spectrogram. Specifically, a window size of $1024$ samples with $50\%$ overlap is used to generate each spectrogram; then, the mean and standard deviation are taken from each frequency bin to produce a feature vector. Similarly, the second feature extraction technique uses Mel-frequency cepstral coefficients (MFCCs) to transform the spectrogram into a Mel scale before extracting features from each frequency bin using the mean and standard deviation. Examples of FFT and MFCC spectrograms are shown in the middle and bottom rows of Fig.~\ref{fig:example_extraction}.

\subsubsection{Machine Learning Model}
Two supervised ML models, a decision tree and a random forest [\onlinecite[Sec.~18]{murphy2022probabilistic}], are trained and evaluated using the AudioMNIST dataset. These models are prone to overfitting but provide simple and explainable decision-based predictions to classify audio samples. Decision trees (DTs) divide the input data by evaluating features, such as frequency, amplitude, or other acoustic characteristics, at each node, creating increasingly specific subsets of the data. This process continues until the data is grouped to minimize variation in the target at the leaf nodes. A random forest (RF) consists of multiple decision trees, each trained on different random subsets of the acoustic data, with each tree making an independent prediction. The final output of an RF is determined by combining the predictions of all trees, often through averaging or voting. This ensemble approach enhances accuracy and reduces the risk of overfitting to particular sound characteristics or noise patterns. At each split in a decision tree, features are selected based on their ability to maximize information gain, typically using measures like entropy, which helps ensure the tree captures the most meaningful acoustic distinctions between sound classes or patterns. 

\subsubsection{Training a model}
Audio clips for each spoken number are divided into even subsets, with $80\%$ of samples used for training and $20\%$ for testing. Clips from each speaker are selected at random. Features are extracted from each clip using the two methods described previously. Training and testing data are normalized and standardized using the feature-wise mean and standard deviation from the training data. Hyperparameters for the decision tree and random forest are chosen using Bayesian Optimization provided in Optuna \cite{akiba2019optuna}. The DT hyperparameter space includes a maximum depth ranging from $4$ to $100$ branches, metrics for split quality (e.g., Gini index, entropy, log loss), and strategies to choose the split at each node (i.e., best or random). The RF hyperparameter space additionally includes the number of estimators ranging from $2$ to $50$. Models are trained with 3-fold cross-validation on the training data [\onlinecite[Sec.~4.5]{murphy2022probabilistic}]. The model with the highest validation accuracy is chosen as the best model. This process is repeated for $20$ instances with varying hyperparameters for each model and feature choice.

\subsubsection{Model Evaluation}
Overall, the DT and RF model performance is shown in Table \ref{acousticclasstable}. The MFCC feature representation has considerably greater performance than the FFT feature representation, and RF models outperformed DT models with both feature representations. The RF model with MFCC feature representation had the highest overall performance. 

\begin{table}
\centering
\begin{tabular}{|l|cccc|}
\hline
Metrics & \multicolumn{2}{c|}{FFT} & \multicolumn{2}{c|}{MFCC} \\ \cline{2-5} 
                         & DT       & RF       & DT       & RF       \\ \hline
Accuracy                 & 0.75          & 0.93          & 0.87          & \textbf{0.99}          \\ 
Precision                & 0.75          & 0.93          & 0.87          & \textbf{0.99}          \\ 
Recall                   & 0.75          & 0.93          & 0.87          & \textbf{0.99}         \\ 
F1 Score                 & 0.75          & 0.93          & 0.87          & \textbf{0.99}          \\ \hline
\end{tabular}
\caption{Model performances for each feature representation.}
\label{acousticclasstable}
\end{table}

Classification performance is visualized using a confusion matrix, as shown in Tables\ \ref{confusionmatrix_ac_fft} and \ref{confusionmatrix_ac_mfcc}. Diagonals and off-diagonals represent the frequency of correct and incorrect classifications, respectively.
The sum of each row indicates the true number of samples for the class, while the sum of each column indicates the number of times a class label is predicted.
Confusion matrices are useful for identifying where a model may have prediction biases.
For example, the RF trained with FFT features (Table~\ref{confusionmatrix_ac_fft}) has many off-diagonal values when predicting class ``one''.
In this instance, the model predicts numbers ``zero'', ``two'', ``four'', or ``nine'' as number ``one''. Analyzing rows of the confusion matrix shows the model has difficulty with the spoken digits ``zero'' and ``two''. Conversely, many off-diagonal values are nearly zero for the RF trained with MFCC features, indicating that the model has near-perfect accuracy, precision, and recall. Although the confusion matrix identifies inaccuracies in model prediction, it does not provide specific examples or a deeper look into why these errors occur. To address such performance issues, detailed inspection of data examples from classes with low accuracies is necessary to understand why a model struggles.

\begin{table}[t!]
\centering
\begin{tabular}{|cc|cccccccccc|}
\hline
&\multicolumn{11}{c|}{\textbf{Predicted Label}} \\ 
\multirow{10}{*}{\rotatebox{90}{\textbf{True Label}}} &\multicolumn{1}{c}{}& \textbf{0} & \textbf{1}& \textbf{2} & \textbf{3} & \textbf{4} & \textbf{5} & \textbf{6} & \textbf{7} & \textbf{8} & \textbf{9} \\
\cline{3-12}
&\textbf{0} &  516 & 5  & 21  & 7  & 4  &  2 & 0  &  31 & 4& 6  \\
%\cline{2-12}
&\textbf{1} & 1  & 569  & 0  & 0  & 5  & 2  & 0  & 0  & 0 & 22\\
%\cline{2-12}
&\textbf{2} & 26  & 12  & 536  & 17  & 5  & 0  & 0  & 2  & 7 & 0\\
%\cline{2-12}
&\textbf{3} & 6  & 0  & 6  & 551  & 0  & 2  & 2  & 0  & 20 & 2\\
%\cline{2-12}
&\textbf{4} &  3 & 21  & 0  & 0  & 574  &  5 & 0  & 2  & 0 & 2\\
%\cline{2-12}
&\textbf{5} &  1 &  4 & 1  & 1  & 6  & 574  & 0  & 0  & 0 & 16\\
%\cline{2-12}
&\textbf{6} & 0  &  0 &  0 &  3 & 0  & 2  & 600  & 1  & 14&0  \\
%\cline{2-12}
&\textbf{7} & 25  & 0  & 3  & 4  & 0  & 1  & 2  & 548  & 1& 5 \\
%\cline{2-12}
&\textbf{8} & 0  & 0  & 4  & 16  & 0  & 0  & 3  & 2  & 557& 1 \\
%\cline{2-12}
&\textbf{9} &  2 &  45 &  1 &  0 &  1 &  8 &  0 & 0&  0&552 \\
\hline
\end{tabular}
\caption{Confusion matrices for the random forest trained with FFT features.}
\label{confusionmatrix_ac_fft}
\end{table}

\begin{table}[t!]
\centering
\begin{tabular}{|cc|cccccccccc|}
\hline
&\multicolumn{11}{c|}{\textbf{Predicted Label}} \\ 
\multirow{10}{*}{\rotatebox{90}{\textbf{True Label}}} &\multicolumn{1}{c}{}& \textbf{0} & \textbf{1}& \textbf{2} & \textbf{3} & \textbf{4} & \textbf{5} & \textbf{6} & \textbf{7} & \textbf{8} & \textbf{9} \\
\cline{3-12}
&\textbf{0} &  579 & 3  & 7  & 4  & 0  & 0 & 0  &  2 & 0& 1  \\
%\cline{2-12}
&\textbf{1} & 2  & 590  & 0  & 0  & 0  & 0  & 0  & 0  & 0 & 7\\
%\cline{2-12}
&\textbf{2} & 3  & 0  & 596  & 2  & 0  & 0  & 0  & 1  & 2 & 1\\
%\cline{2-12}
&\textbf{3} & 7  & 0  & 4  & 574  & 0  & 0  & 0  & 0  & 4 & 0\\
%\cline{2-12}
&\textbf{4} &  2 & 2  & 0  & 0  & 602  &  0 & 0  & 1  & 0 & 0\\
%\cline{2-12}
&\textbf{5} &  0 &  1 & 0  & 0  & 4  & 597  & 0  & 1  & 0 & 0\\
%\cline{2-12}
&\textbf{6} & 0  &  0 &  0 &  0 & 0  & 0  & 617  & 3  & 0&0  \\
%\cline{2-12}
&\textbf{7} & 0  & 0  & 1  & 0  & 0  & 1  & 0  & 586  & 0& 1 \\
%\cline{2-12}
&\textbf{8} & 1 & 0  & 0  & 2  & 0  & 1  & 1  & 1  & 577& 0 \\
%\cline{2-12}
&\textbf{9} &  0 &  9 &  0 &  0 &  0 &  0 &  0 & 1&  0&599 \\
\hline
\end{tabular}
\caption{Confusion matrices for the random forest trained with Mel-frequency cepstral coefficients (MFCC) features.}
\label{confusionmatrix_ac_mfcc}
\end{table}

\subsection{Acoustic data exploration with unsupervised ML}

In cases involving large acoustic data sets—for instance, those generated by continuously recording sensor arrays—the sheer volume of data can make manual analysis impractical.
Given time and resource constraints, sacrificing every audio snippet for meaningful insights may be unfeasible.
To overcome this challenge, unsupervised ML techniques can be employed to analyze and categorize the data systematically.
Unsupervised ML involves algorithms and models that learn patterns and structures from data without explicit supervision or labeled target outputs.\cite{bishop2006pattern, murphy2022probabilistic}
Specifically, clustering algorithms seek to discover similar examples within the data and are used for data mining and exploratory data analysis.\cite{bishop2006pattern, murphy2022probabilistic}
Clustering results can then guide further analysis, such as targeted review of specific segments or features of interest, or removal of unwanted data types.

Many clustering algorithms perform more effectively when working with lower-dimensional data.\cite{aggarwal2014data}
This is particularly relevant for acoustic datasets, which typically contain thousands or millions of features when represented as time series, spectrograms, scalograms, or energy envelopes.
The high dimensionality of these representations—whether discrete samples in a time series or time-frequency bins in a spectrogram—presents computational challenges for standard clustering approaches.
To address this limitation, we present a workflow that combines autoencoders (deep neural networks specialized in dimensionality reduction) with clustering algorithms.
A code implementation of this workflow can be found in Chapter~3.5 in the AcousticsML repository.
The dimensionality reduction of autoencoders has been paired with both supervised and unsupervised ML workflows in a variety of acoustic\cite{ozanich2021deep, linhardt2022cost, desalvio2023blind, guerrero2023acoustic, jedrusiak2024interdisciplinary, gibb2024interpretable, liu2025modeinformed, zhang2025exploring} and seismic\cite{mousavi2019unsupervised, snover2021deep, jenkins2021unsupervised, chien2023automatic} settings.

\begin{figure}
    \centering
    \includegraphics[width=.99\linewidth]{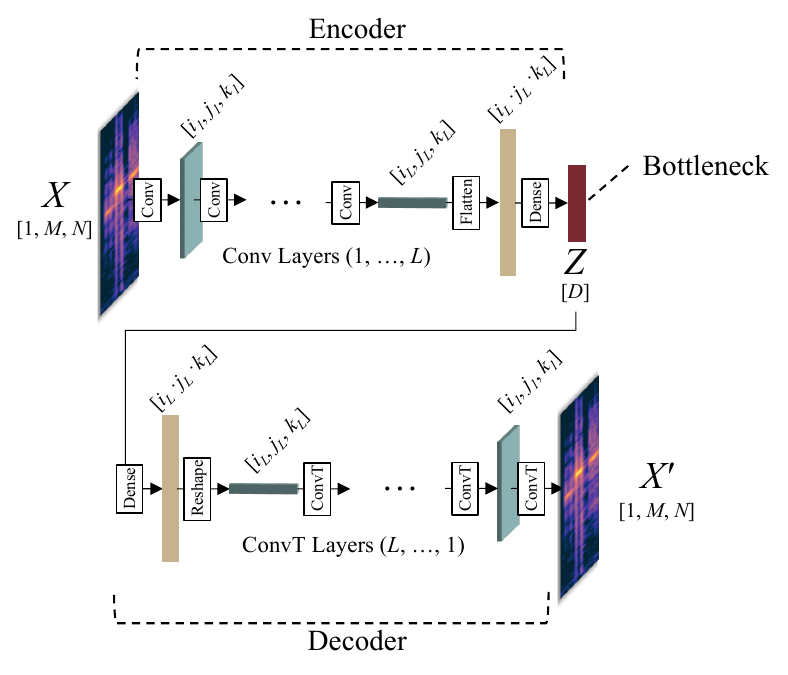}
    \caption{Architecture of an autoencoder network. The encoder maps input data $X$ to a latent representation $Z$, and the decoder reconstructs the input from the latent representation. Embeddings in $Z$ are a reduced dimensionality representation of the input data and can be used for clustering or other tasks.}
    \label{fig:ae_network}
\end{figure}

\subsubsection{Dimensionality reduction with autoencoders}\label{dimred_aec}

A sample from an acoustic data set can be represented as a vector $\mathbf{x} = [x_1,\dotsc,x_N]^\mathsf{T} \in \mathbb{R}^{N\times 1}$, where each feature corresponds to an element $x_n$ in the vector $\mathbf{x}$ which describes a point in $N$-dimensional space.
Directly clustering high-dimensional data is vulnerable to the ``curse of dimensionality": \cite{aggarwal2014data, murphy2022probabilistic} as the dimensionality of the input data increases linearly, the number of data points required to maintain sufficient sampling density increases exponentially.
Additionally, clustering algorithms can give less meaningful results as dimensionality increases, making clustering in high dimensions challenging and unreliable.\cite{aggarwal2014data}

\begin{figure*}
    \centering
    \includegraphics[width=.99\linewidth]{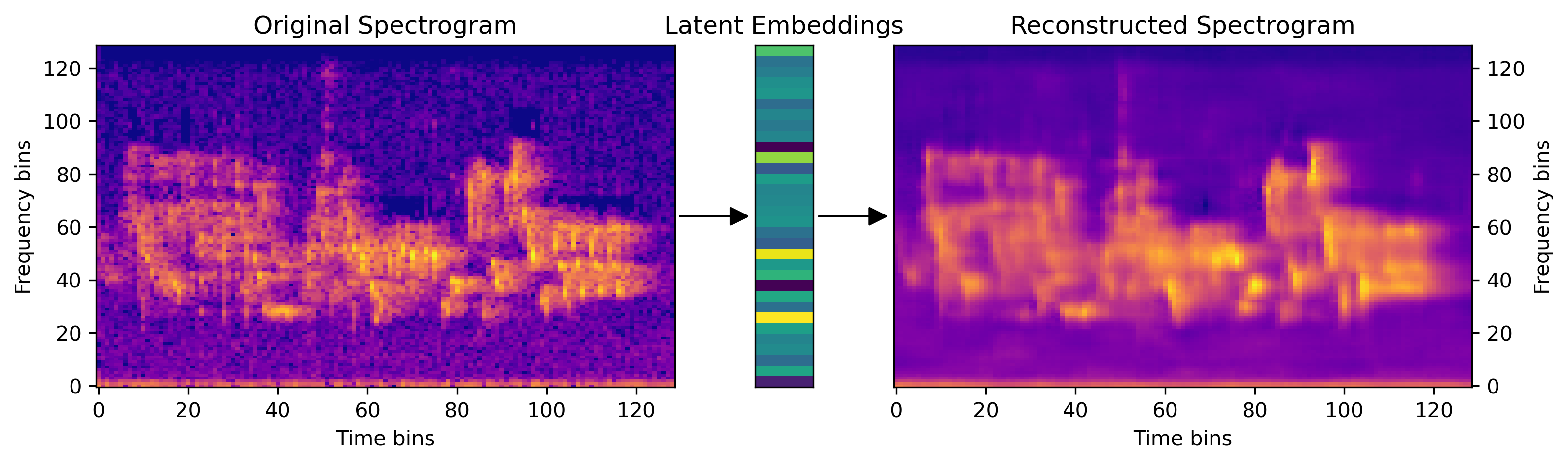}
    \caption{Spectrogram latent embedding example.} A spectrogram (left) is provided to a convolutional autoencoder trained on birdsong. The 32-dimensional latent embeddings (center) contain the salient features required to produce the reconstructed spectrogram (right).
    \label{fig:ae_inout}
\end{figure*}

A popular approach is principal component analysis (PCA), which projects higher-dimensional data into a lower-dimensional space [\onlinecite[Sec.~20.1]{murphy2022probabilistic}].
However, PCA is a linear method and may not be effective for data with complex, nonlinear structures.
An alternative model that can capture nonlinear relations is an autoencoder, a neural network that learns to encode data into a latent, lower-dimensional representation.\cite{murphy2022probabilistic}
A typical autoencoder architecture is shown in Fig.~\ref{fig:ae_network} and consists of three components: an \textit{encoder}, a \textit{bottleneck}, and a \textit{decoder}\cite{murphy2022probabilistic}.
First, the encoder maps input data, like spectrograms, from a data space $X$ into a latent feature space $Z$ by $f_{\theta}: X \to Z$, where $\theta$ are the neural network parameters.
Next, the decoder attempts to reconstruct $X$ from $Z$ by ${g_{\theta}: Z \to X'}$.
An entire forward pass through the autoencoder is represented as
\begin{equation}
    F_\theta: X \to Z \to X',\quad F_\theta = g_\theta \circ f_\theta.
    \label{eq:autoencoder}
\end{equation}
The autoencoder is trained by iteratively updating $\theta$ through backpropagation (Ref.~\onlinecite[Sec.~6.5]{goodfellow2016deep}) to minimize a loss function defined as the reconstruction error between $X$ and $X'$, e.g., the mean squared error ${\textrm{MSE}(X - X')}$.
In minimizing the error, the autoencoder learns the salient features of $X$ and accurately embeds them in $Z$, enabling subsequent tasks like clustering to be performed in the lower-dimensional latent space. A successor to autoencoders, the variational autoencoder (VAE), enables data generation from the latent space and is introduced in Sec.~\ref{ssec:vae}.

An example of an autoencoder applied to a spectrogram is shown in Fig.~\ref{fig:ae_inout}.
The spectrogram contains 129 time bins and 129 frequency bins, producing 16,641 features.
The encoder maps the spectrogram to a latent representation with 32 features, and the decoder reconstructs the spectrogram from the latent representation back to the original 16,641 features.

\subsubsection{Clustering}

After reducing the data dimensionality with the autoencoder, clustering algorithms can be applied to the latent space to group similar data points.
One of the most common clustering algorithms is \textit{k}-means, which partitions data into $K$ clusters by minimizing the sum of squared distances between data points and their cluster centroids [\onlinecite[Sec.~21.3]{murphy2022probabilistic}].
However, \textit{k}-means makes assumptions about the data, such as isotropic clusters and balanced cluster populations, which may not hold in practice.
A more general approach is to model clusters as a mixture of $K$ multivariate Gaussian distributions.
GMMs can capture anisotropic and imbalanced clusters and yield probabilities that each point belongs to a particular cluster, enabling a more in-depth analysis of the clustering results [\onlinecite[Sec.~21.4]{murphy2022probabilistic}].

Determining the optimal number of clusters, $K$, is a challenging problem in unsupervised machine learning.
Furthermore, when autoencoders are used in conjunction with clustering in the latent space, care must be taken to ensure that clustering results map to meaningful features in the original data space.
The choice of $K$ significantly affects the clustering results, and selecting an inappropriate number of clusters can lead to suboptimal or misleading results.
The choice for $K$ should be evaluated through both qualitative inspection and quantitative metrics.
Clustering results are qualitatively evaluated for similarity of data points to their respective cluster centroids and to data points within a cluster, and should be done in both the latent and data spaces.
Quantitative evaluation can be performed using metrics such as the gap statistic\cite{tibshirani2001estimating} and silhouette scores [\onlinecite[Sec.~21.3.7.3]{murphy2022probabilistic}], which measure the compactness and separation of clusters.

\subsection{Generative modeling for spatial audio and room acoustics}

Generative models aim to model the distribution of data and use ML models to generate more synthetic data that seems to be from the same distribution. A typical ML method for generative modeling is GMM, assuming the data is a mixture of a Gaussian distribution and can sample from each distribution to generate a new data sample.

\subsubsection{Generative Adversarial Networks for room acoustics}

Chapter 5.2 presents a Jupyter Notebook using a generative adversarial network to generate room impulse responses.

Generating Room Impulse Responses (RIRs) is an important research topic due to their role in capturing the acoustic characteristics of an environment. Reverberant sound, which reflects the layout and materials of the environment, is crucial in human spatial awareness. However, in applications such as automatic speech recognition, this reverberation can act as noise, degrading system performance by masking speech clarity or introducing distortions. RIRs serve as a fundamental representation of sound propagation in a room, assuming the system behaves as linear and time-invariant. Despite their significance, RIRs are typically challenging to measure directly, as they require specialized setups, including loudspeakers to emit a sine sweep and microphones to record the response at various locations. Post-processing is necessary to extract the RIR, adding complexity to their practical use. 

Generating RIRs through ML methods offers an attractive solution to overcome these challenges and improve the robustness of acoustic models in real-world environments. 

In this example, we use the BUT ReverbDB dataset~\cite{szoke2019building}, which contains a dataset of real room impulse responses (RIRs). Following IR-GAN architecture~\cite{ratnarajah21_interspeech}, the generator comprises 5 layers of 1D deconvolution, and the discriminator consists of 5 layers of 1D convolution. Based on a standard GAN training setup, the model can generate plausible room impulse responses.

\subsubsection{Personalized HRTF modeling with Implicit Neural Representations}

Chapter 5.3 of the AcousticsML repository presents a Jupyter Notebook on using implicit neural representations to model HRTFs across datasets.

Spatial audio plays a pivotal role in creating immersive experiences through headphones or VR headsets, allowing users to perceive the direction and distance of the sound. By incorporating individual uniqueness into auditory perception, spatial audio rendering can significantly enhance the sense of immersion. A key aspect of this task is to predict personalized head-related transfer functions (HRTFs), which describe the spatial filtering effects of human geometry for accurate sound localization.

Due to the resource-intensive nature of HRTF measurements, existing databases have a limited number of subjects, which poses challenges for data-intensive machine learning models. HRTFs are inherently high-dimensional, encompassing numerous spatial locations and frequency bins per subject.

Human geometry input can be in several formats, including anthropometric measurements, ear images, or a scanned head mesh, arranged from lower dimension to higher complexity. Usually, these data are fed into an encoder to map to a latent space, and then the HRTF is decoded from the latent space as the output. Typical ML models include autoencoders, variational autoencoders, and generative adversarial networks (GAN)~\cite{fantini2025survey}.

Personalized HRTF modeling involves two tasks: interpolation and personalization. As measurement is time-consuming, the number of locations is usually limited. With generative modeling, getting the HRTF at arbitrary locations would be ideal. These include generating the HRTF, given the existing measured HRTFs as conditions, or modeling the whole distribution of the HRTFs for various subjects. 

\begin{figure}
    \centering
    \includegraphics[width=0.9 \linewidth]{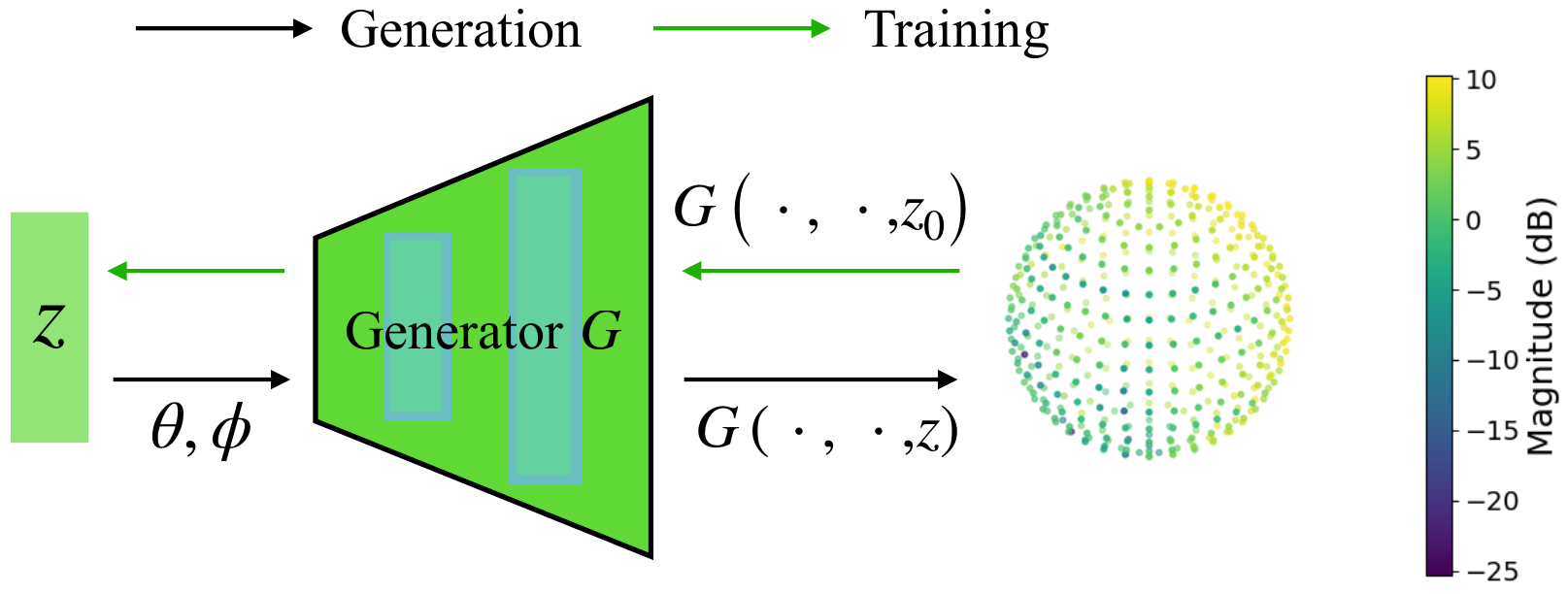}
    \caption{HRTF modeling with implicit neural representation. The colormap corresponds to the magnitude in the logarithmic (dB) domain at the 2 kHz frequency bin.
    } 
    \label{fig:HRTF}
\end{figure}

In the Jupyter Notebook, we present the use of INRs to interpolate the HRTFs~\cite{Zhang2023HRTFfield}. The HRTF of each person at azimuth $\theta$ and elevation angle $\phi$ is modeled with the output of the generator $G(\theta, \phi, z)$, where the latent vector $z$  represents the personalized HRTF of each person. The training and generation process is illustrated in Fig.~\ref{fig:HRTF}.

We use the HUTUBS dataset~\cite{brinkmann2019cross} and build the INR with a 2-layer multi-layer perceptron with 2048 nodes in each layer. The model is trained in an autodecoder fashion, where the latent vector $z$ is first assumed at the origin and then updated with the negative gradient. 
\begin{equation}
\mathbf{z} = \mathbf{z}_0 - \nabla_{\mathbf{z}_0} \mathcal{L}_{\mathrm{MSE}}\left(\mathbf{x}, G\left(~\boldsymbol{\cdot}, ~\boldsymbol{\cdot}, \mathbf{z}_0\right)\right).
\label{eq:z0}
\end{equation}
With the new $z$, the generator G is updated with the $\ell_2$ distance between the generated HRTF and the ground truth.
\begin{equation}
\mathcal{L}=\mathcal{L}_{\mathrm{MSE}}\left(\mathbf{x}, G\left(~\boldsymbol{\cdot}, ~\boldsymbol{\cdot}, \mathbf{z}\right)\right).
\label{eq:update}
\end{equation}

\subsection{Physics-informed neural networks (PINN)}
Chapters 5.4 and 5.5 include two Jupyter Notebooks to solve forward and inverse problems in acoustics using PINNs. 
These notebooks use a finite difference solver (included in the repository) to generate data and ground truth solutions. 
\subsubsection{Problem formulation}
In Chapter 5.4, the goal is to solve the time-domain wave equation,
\begin{equation}
    \nabla p(\mathbf{r},t) - \frac{1}{c^2(\mathbf{r})} \frac{\partial^2 p(\mathbf{r},t)}{\partial t^2} = 0,
    \label{eq:wave_eq}
\end{equation}
given a known wave speed, $c(\mathbf{r})$. 
It is assumed that the wavefield at some initial time steps,
\begin{equation}
    p_0 = p(\mathbf{r},t) \quad \text{for} \quad t \leq t_0,
\end{equation}
where $t_0$ is an early time for which the wave has not yet propagated far, is known. These snapshots contain the source position, shape, and the early wave propagation.
The domain boundaries are considered absorptive so that there are no reflected waves. With this information, the goal is to compute the wave field $p(\mathbf{r},t)$.

\subsubsection{Loss function}
The neural network used to represent the wave field is denoted $\hat{p}(\mathbf{r},t;\boldsymbol{\theta}_p)$. The network's input is the spatiotemporal coordinates, $(\mathbf{r},t)$, and its output is the computed wavefield. The network parameters  $\boldsymbol{\theta}_p$ are tuned during training to minimize a loss function composed by the weighted sum of two terms,
\begin{equation}
    \mathcal{L} = \lambda_\text{pde}\mathcal{L}_\text{pde} + \lambda_\text{ic}\mathcal{L}_\text{ic}.
    \label{eq:loss_function_pinn}
\end{equation}
The first one is the physics term, which constrains the network's output to satisfy the wave equation 
\begin{equation}
    \mathcal{L}_\text{pde} \!= \!\frac{1}{n_\text{pde}} \sum_{i=1}^{n_\text{pde}} \!\left \| \nabla^2 \hat{p}(\mathbf{r}_i,t_i;\boldsymbol{\theta}_p) - \frac{1}{c^2} \frac{\partial^2 \hat{p}(\mathbf{r}_i,t_i;\boldsymbol{\theta}_p)}{\partial t^2} \right\|^2,  
    \label{eq:loss_pde}
\end{equation}
where $(\mathbf{r}_i,t_i), i=1,\dots,n_\text{pde}$ are stochastically sampled over the spatio-temporal domain during training, with $n_\text{pde}$ user-chosen. The second term imposes the initial condition, 
\begin{equation}
    \mathcal{L}_\text{ic} = \frac{1}{n_\text{ic}} \sum_{j=1}^{n_\text{ic}} \left \| \hat{p}(\mathbf{r}_j,t_j;\boldsymbol{\theta}_p) - p_0(\mathbf{r}_j,t_j) \right\| 
\end{equation}
where $t_j\leq t_0$ and $(\mathbf{r}_i, t_j), j=1,\dots,n_\text{ic}$ are points sampled at early times.

This is a `soft-constrained' formulation, meaning that the constraints guide the training but are not enforced in a hard way. Such soft-constrained PINNs are easy to formulate, but the learned function might not satisfy the conditions exactly. Constraints can also be imposed using the neural network as part of a solution ansatz. 

\begin{figure*}
    \centering
    \includegraphics[width=0.9\linewidth]{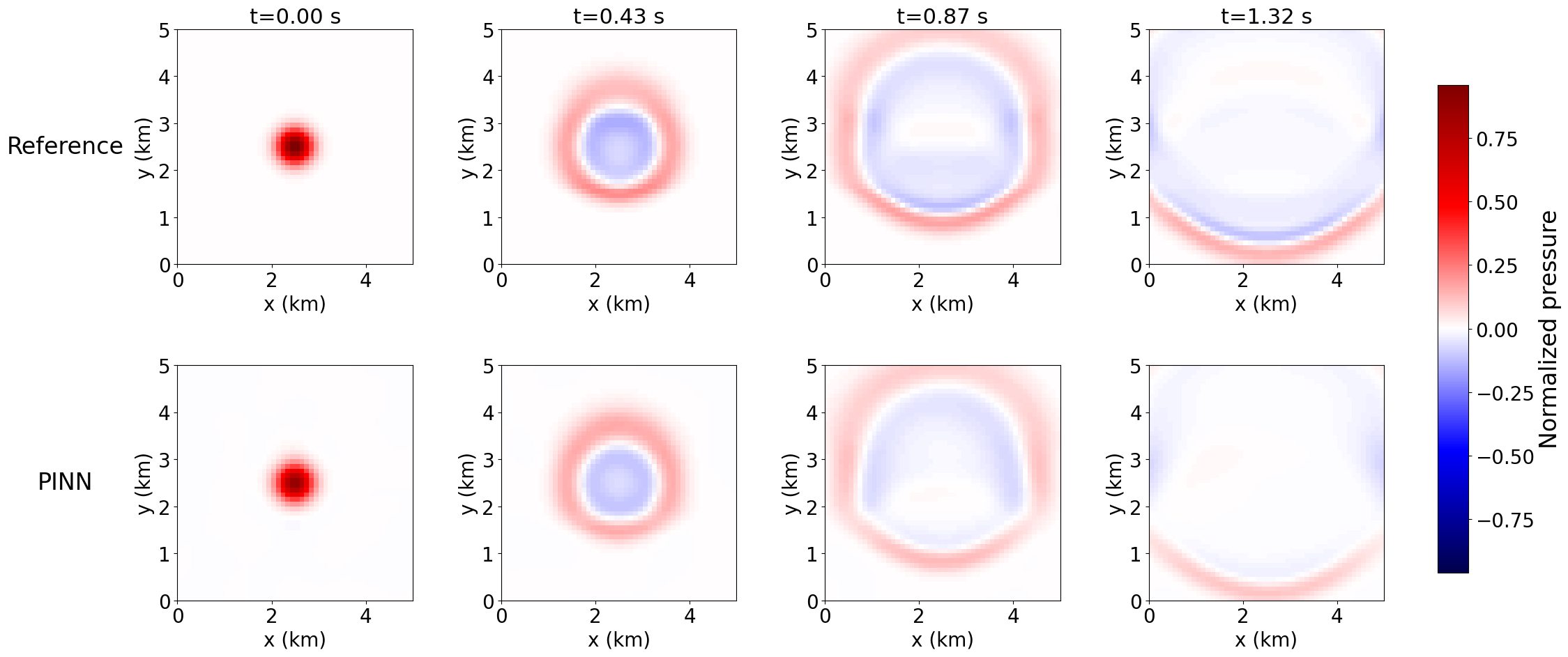}
    \caption{Forward solution of the wave equation in a 2D domain with a stratified wave speed and a Gaussian pulse as the initial condition. Top row: reference solution. Bottom row: PINN estimation.}
    \label{fig:pinn_forward}
\end{figure*}

Figure \ref{fig:pinn_forward} presents an example of using PINNs for solving the wave equation in a medium with a stratified wave speed. The initial condition is a Gaussian pulse in the center of the domain. A PINN is trained to minimize the loss of Eq. \ref{eq:loss_function_pinn}. After training, the output of the PINN approximates the wavefield, $p(\mathbf{r},t)$. This example corresponds to Chapter 5.4. 

\subsubsection{Balancing the loss function}
The weights $\lambda_\text{pde}$ and $\lambda_\text{ic}$ balance the two terms in the loss function. The choice of the weights is delicate, and manually finding weights that properly balance the loss terms can be difficult. In the notebooks, we implement an annealing algorithm\cite{wang2021understanding} to automatically choose $\lambda_\text{pde}$ and $\lambda_\text{ic}$. The weights are chosen based on the gradient of each loss term with respect to the network parameters to prevent an imbalance of the back-propagated gradients during training. 

\subsubsection{Network architecture}
For the PINN architecture, a fully connected network with hyperbolic tangent activation functions, three layers, and 64 units per layer is chosen. This type of simple architecture is often used for PINNs, and it is convenient for this example.  

\subsubsection{Fourier features}
Neural networks (not only PINNs) often suffer from \textit{spectral bias}, i.e., they struggle to learn the high-frequency content of functions. High frequencies can be associated, for example, with function discontinuities or jumps, such as edges in an image. 
This is particularly relevant in the case of PINNs for acoustics because acoustic sources tend to contain a broad spectrum of frequencies. The use of Fourier features has been proposed as a way of alleviating the spectral bias of PINNs in multiscale problems.\cite{wang2021eigenvector}
A Fourier mapping of a network with inputs $\mathbf{r} \in \mathbb{R}^{n_\text{in}}$ can be computed as
\begin{equation}
   {\bm \gamma}(\mathbf{r}) = \begin{bmatrix} \cos \mathbf{(Br)} \\ \sin \mathbf{(Br)} \end{bmatrix}
\end{equation}
where $\mathbf{B}\in \mathbb{R}^{n_\text{ff}\times n_\text{in}}$, and $n_\text{ff}$ is the number of Fourier features in the mapping. The entries of $\mathbf{B}$ are sampled from a normal distribution with variance $\sigma^2$. It has been shown, however, that the choice of $\sigma^2$ is not straightforward since it depends on the frequency content of the function to be approximated,\cite{wang2021eigenvector}, and different $\sigma^2$ should be chosen for the spatial coordinates and the temporal ones. In the notebook, we change the formulation slightly and compute the Fourier mapping as a cosine transform with an offset,
\begin{equation}
    {\bm \gamma}(\mathbf{r}) = \cos \mathbf{(Br + b)},
\end{equation}
where $\mathbf{B} \in \mathbb{R}^{n_\text{ff}\times n_\text{in}}$  and $\mathbf{b}\in \mathbb{R}^{n_\text{ff}}$ are treated as network parameters that are learned during training.
This $\bm\gamma$ is then input to the first layer of the neural network.

\subsubsection{Causal training}
A limitation of the original PINNs formulation is their inability to respect the spatio-temporal causality of the modeled physical system.\cite{wang2024respecting} PINNs modeling transient phenomena are often biased toward minimizing the residual at later times before learning the initial condition. This causes the PINN to get stuck in a local minimum and learn an erroneous solution. 

Different strategies have been proposed to enforce causality.\cite{mattey2022novel,wang2024respecting} 
In the notebook, we implement a simple curriculum learning scheme
where the PINN is trained over successively increasing time segments. In other words, we control the collocation points in Eq.~\eqref{eq:loss_function_pinn} during training so that we only feed $t_{i+1}$ to the PINN after minimizing the loss at $t$.

\subsubsection{Inverse estimation problem}
The second notebook, Chapter 5.5, presents an example of PINNs solving an inverse estimation problem. In this case, the wave speed $c(\mathbf{r})$ and initial condition $p_0$ are unknown, and the information available is discrete observations of the wave field at several locations, $p_\text{obs} = p(\mathbf{r}_j,t_j)$, $j=1,\dots,n_\text{obs}$. The inverse problem aims to estimate the wave speed $c(\mathbf{r})$ from the observations. 

For solving the inverse problem we train two neural networks: $\hat{p}(\mathbf{r},t;\boldsymbol{\theta}_p)$, which approximates the wave field, and $\hat{c}(\mathbf{r};\boldsymbol{\theta}_c)$, which approximates the wave speed. Both networks are trained simultaneously with a single loss function: 
\begin{equation}
    \mathcal{L} = \lambda_\text{pde}\mathcal{L}_\text{pde}(\boldsymbol{\theta}_p, \boldsymbol{\theta}_c) + \lambda_\text{obs}\mathcal{L}_\text{obs}(\boldsymbol{\theta}_p), 
    \label{eq:loss_function_inv}
\end{equation}
where 
\begin{equation}
    \mathcal{L}_\text{pde} \!= \!\frac{1}{n_\text{pde}}\! \sum_{i=1}^{n_\text{pde}} \!\left \| \nabla^2 \hat{p}(\mathbf{r}_i,t_i;\boldsymbol{\theta}_p) - \frac{1}{\hat{c}(\mathbf{r}_i; \boldsymbol{\theta}_c)^2} \!\!\!\!\!\!\!\frac{\partial^2 \hat{p}(\mathbf{r}_i,t_i;\boldsymbol{\theta}_p)}{\partial t^2} \right\|^2\!,  
\end{equation}
and
\begin{equation}
    \mathcal{L}_\text{obs} = \frac{1}{n_\text{obs}} \sum_{j=1}^{n_\text{obs}} \left \| \hat{p}(\mathbf{r}_j,t_j;\boldsymbol{\theta}_p) - p_\text{obs}(\mathbf{r}_j,t_j) \right\|.
\end{equation}
The PDE loss term links the two networks, as the wave equation contains both $p(\mathbf{r},t)$ and $c(\mathbf{r})$.
Once trained, $\hat{c}(\mathbf{r})$ can be queried at any point $\mathbf{r}$ to obtain an estimate of the wave speed. 

\section{Conclusion and Future Perspectives}
We present an ML review that provides an in-depth discussion of applications on our open-source GitHub page. We demonstrate typical approaches for applying ML to acoustic research, focusing on applications such as sound classification, generative modeling for synthetic data, and physics-informed ML. The notebooks and the paper focus on a few relevant topics, emphasizing the benefits of applying ML and evaluating performance. Central to ML, trained models must generalize well on unobserved data---the test data.

ML methods can find structure in the data and learn low-dimensional representations of complex physical phenomena, like wave propagation. A current trend, likely to continue, is the learning of tractable surrogate models to accelerate simulations, processing, and estimation tasks.

The integration of physical constraints into ML models has improved their accuracy, efficiency, interpretability, and generalizability. We foresee a closer integration of physics and domain-specific knowledge into ML models, for example, by designing custom architectures that inherently comply with the physics of the problem \cite{chen2018neural}, or the integration of PDE numerical solvers that allow for gradients to flow in the training process \cite{zhu2021general}.

Explainability and interpretability has a central role in the current development of AI. ML methods for scientific discovery, where interpretable mathematical expressions are learned from data, could have an impact on fundamental acoustic research. These include symbolic regression methods \cite{wang2023scientific}, and the use of small but efficient neural networks like Kolmogorov-Arnold networks \cite{liu2024kan}.

\begin{acknowledgments}
Part of the content of this manuscript was presented at the tutorial session of the ASA 187th Meeting, held Online, Nov 20, 2024~\cite{gerstoft2024tutorial}. P.G. thanks the support from the Office of Naval Research, N000142412016. W.F.J. thanks the support from the Office of Naval Research, N00014-24-1-2401.
\end{acknowledgments}

\begin{contributions}
R.A.M. prepared Figures 5-8. Y.Z. prepared Figures 2 and 11. S.A.V. prepared Figures 1, 3, and 12. W.F.J. prepared Figures 9 and 10. P.G. prepared Figure 4. All authors contributed equally to the conception, code development, original draft writing, review, and approval of the submitted manuscript.
\end{contributions}

\begin{competinginterests}
The authors declare no competing interests.
\end{competinginterests}

\begin{dataavailability}
    Code for the manuscript can be found at the AcousticsML repository at \url{https://github.com/RAMshades/AcousticsML}.
\end{dataavailability}

\bibliographystyle{unsrtabbrv1} % <-- Note that a custom jabbrv-
\bibliography{MLMarch5.bib}

\end{document}